\documentclass[journal=ancac3,manuscript=article]{achemso}

\usepackage[version=3]{mhchem} 

\usepackage{graphicx}     
\usepackage{stackengine}
\DeclareTextFontCommand{\helvet}{\fontfamily{phv}\selectfont}

\author{Delphine Dessaux}
\author{J\'er\^ome Math\'e}
\author{Rosa Ramirez}
\author{Nathalie Basdevant}
\email{nathalie.basdevant@univ-evry.fr}
\affiliation{Universit\'e Paris-‐Saclay, CNRS, Univ Evry, LAMBE, 91025, \'Evry‐-Courcouronnes, France}

\title{Current rectification and ionic selectivity of $\alpha$-hemolysin: Coarse-Grained Molecular Dynamics simulations}

\abbreviations{MD, CG, AA, $\alpha$HL}

\begin{document}


\begin{abstract}

In order to understand the physical processes of nanopore experiments at the molecular level, microscopic information from molecular dynamics is greatly needed. Coarse-grained models are a good alternative to classical all-atom models since they allow longer simulations and application of lower electric potentials, closer to the experimental ones. We performed coarse-grained molecular dynamics of the ionic transport through the $\alpha$-hemolysin protein nanopore, inserted into a lipid bilayer surrounded by solvent and ions. For this purpose, we used the MARTINI coarse-grained force field and its polarizable water solvent (PW). Moreover, the electric potential difference applied experimentally was mimicked by the application of an electric field to the system. 
We present, in this study, the results of 1.5 microsecond long-molecular dynamics simulations of twelve different systems for which different charged amino acids were neutralized, each of them in the presence of nine different electric fields ranging between $\pm 0.04$~V.nm$^{-1}$ (a total of around 100 simulations). We were able to observe several specific features of this pore, current asymmetry and anion selectivity, in agreement with previous studies and experiments, and identified the charged amino acids responsible for these current behaviors, therefore validating our coarse-grain approach to study ionic transport through nanopores. We also propose a microscopic explanation of these ionic current features using ionic density maps.

\end{abstract}

\section{Introduction} 
The nanopore technique is a popular method to study the kinetics and energetics of biological phenomena. It consists in applying an electric potential difference to guide a charged polymer through an artificial or protein nanopore, inserted in a solid or lipid membrane, in the presence of a salt solution~\cite{Kasianowicz1996, Dudko2010, Meller2000, Oukhaled2007, Oukhaled2008}. A macromolecule passing through the pore induces a decrease in the ionic current by partially blocking it, depending on the nature of the molecule and the pore characteristics: size, conformation, and structure. The $\alpha$-hemolysin ($\alpha$HL) is a biological protein nanopore widely employed in nanopores experiments. Inserted in a lipid bilayer, it has been used to study DNA or RNA translocation~\cite{Kasianowicz1996, Meller2000}, DNA sequencing or biosensing~\cite{Stoddart2009, Stoddart2010, Deamer2016}, DNA unzipping of individual hairpins~\cite{Mathe2004, Mathe2005, Muzard2010}, and  the translocation of proteins~\cite{Oukhaled2007, DiMarino2015, DiMuccio2019, Bonome2019}, polysaccharides~\cite{Fennouri2018} or synthetic polymers~\cite{Krasilnikov2006, Reiner2010}. 

Several experimental~\cite{Kasianowicz1996,Merzlyak2005,Piguet2014,Payet2015}, theoretical~\cite{Misakian2003} and computational studies~\cite{Noskov2004, Aksimentiev2005, Millar2008, Bhattacharya2011, Gamble2014, Manara2015, Bonome2017, Zhou2020} on the ionic current through $\alpha$HL inserted in a lipid bilayer were already performed and they have shown that this current is not symmetrical for positive and negative applied electric potentials. This effect is called current rectification. It can be explained by the asymmetric charge distribution of the amino acids of the pore which yields an asymmetric electric potential, resulting in an asymmetric crossing of ions under a positive or negative applied voltage difference. 
It should be noted that this effect is specific to each protein pore. Indeed, the rectification observed for other protein nanopores differs from $\alpha$HL, as exemplified by Payet \latin{et al.} for the aerolysin pore~\cite{Payet2015}. 
Zhou \latin{et al.}~\cite{Zhou2020} recently showed with MD simulations of several protein nanopores that the channel geometry and charge distribution affect ion transport.
In addition, several experiments pointed that cation type, temperature, ionic concentration and pH can influence the rectification of $\alpha$HL. For example, Piguet \latin{et al.}~\cite{Piguet2014} showed that the rectification is more important for KCl solution than for LiCl. 
Battacharya \latin{et al.}~\cite{Bhattacharya2011} performed experiments and Molecular Dynamics (MD) simulations with different cation types and found a greater conductance for K$^+$ than for Na$^+$ and Li$^+$, which they associated to the cations ability to screen the pore amino acids charges.
Besides, Payet \latin{et al.}~\cite{Payet2015} conducted experiments showing that, while the ionic current increases with temperature, the current asymmetry decreases. 
Several ionic transport experimental studies~\cite{Menestrina1986, Merzlyak2005, Bonthuis2006} with different KCl concentrations have also proven that conductance increases with concentration whereas asymmetry decreases. 
As for the pH influence, it has been shown both by experiments~\cite{Misakian2003} and all-atom MD simulations~\cite{Bonome2017} that the current asymmetry decreases with low pH. 
In this study, we will focus on the current asymmetry and ionic selectivity of $\alpha$HL. 

\begin{figure}[htb]
	\centering
    \includegraphics[width=0.6\textwidth]{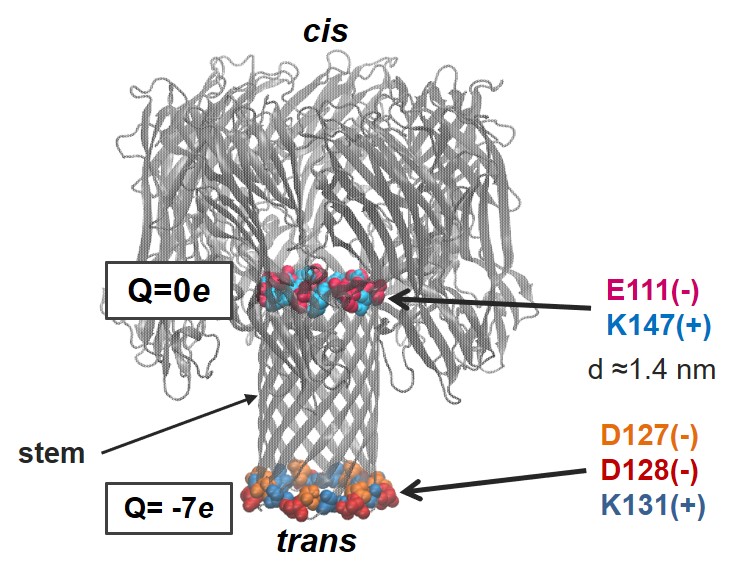}
    \includegraphics[width=0.34\textwidth]{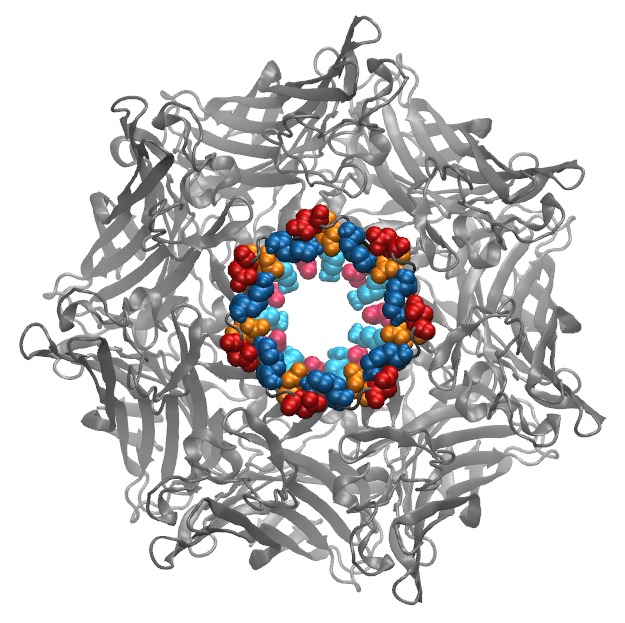}
    \caption{Representation of $\alpha$-hemolysin in cartoon mode. The five charged amino acids in the stem are highlighted in different colors. At the pore constriction (E111 and K147) and at the \latin{trans} side (D127, D128 and K131). On the right, view from the \latin{trans} side.}
    \label{fig:aHL_charges} 
\end{figure}

The $\alpha$HL nanopore structure counts seven identical chains composed of 293 amino acids, and it involves a hydrophobic stem part (mostly composed of beta-sheets), which is embedded into the lipid membrane, and a cap part, facing the \latin{cis} side of the membrane (see Fig.~\ref{fig:aHL_charges}). 
The \latin{trans} part of $\alpha$HL includes three charged amino acids on each of the seven chains: two aspartic acids (D127 and D128) and one lysine (K131), resulting in a total charge of -7e (see Figure~\ref{fig:aHL_charges}). The lateral chains of D127 and K131 point at the inner channel, while D128 are directed towards the membrane. 
Studies involving Poisson-Nernst-Planck (PNP) theory~\cite{Noskov2004}, as well as all-atom MD simulations~\cite{Bhattacharya2011}, revealed that the neutralization of the charged amino acids at the \latin{trans} part induces the disappearance of the current rectification. Moreover, Merzlyak \latin{et al.}~\cite{Merzlyak2005} performed mutations of some charged amino acids in the pore stem which revealed that the residue charges at the \latin{trans} part of the pore affect the rectification.

Furthermore, the $\alpha$HL pore is slightly anion selective. This means that, for all applied voltages through the pore, more anions are crossing over than cations. The charged amino acids inside the protein stem also play a role in this effect. 
In addition to the three charged amino acids at the \latin{trans} part, there are two charged amino acids at the top of the stem, \latin{i.e.} at the pore constriction: one glutamic acid (E111) and one lysine (K147) on each of the seven chains. 
Neutralizing these two amino acids affects the anion selectivity of the pore. Using PNP~\cite{Noskov2004}, these two charged residues were indeed identified to influence the selectivity. Besides, experiments and all-atom MD~\cite{Bhattacharya2011} showed that this anion selectivity was more important with K$^{+}$ cations than with Na$^{+}$ or Li$^{+}$. 
However, both theoretical studies pointed out that neutralization of the charged residues at the \emph{trans} part increased weakly the selectivity. 
Moreover, mutation experiments~\cite{Gu2000,Mohammad2010} revealed that the mutation of E111 and K147 to neutral residues results in cation selectivity. 
Figure~\ref{fig:aHL_charges} represents the $\alpha$HL structure (PDB: 7AHL) and the location of the charged amino acids in the stem.

Here, we choose to study these two effects, rectification and selectivity, performing Coarse-Grained (CG) molecular dynamics simulations, using the MARTINI force field. 
We previously showed that this approach is adapted to study the ionic current through $\alpha$HL inserted into a DPPC bilayer at 1~M KCl concentration~\cite{Basdevant2019}. Indeed, CG-MD simulations allow to simulate the system for longer timescales than All-Atom (AA) simulations and with voltages closer to experimental ones. Though, we showed that there is a scaling factor for ionic conductances: the simulated currents are lower than experimental ones. 
In this study, we aim to identify the role of the charged amino acids in current asymmetry and anion selectivity. Therefore, we calculated ionic currents in the presence of several external electric fields, when specific charged amino acids inside the pore stem were neutralized, performing around 100 different simulations of 1.5 microsecond. We decomposed the ionic current into its anion and cation contribution, in order to evaluate the pore selectivity. We tested many more different combinations of amino acid neutralization compared to previous AA-MD simulations. We were thus able to identify the amino acids responsible for the rectification, at the \latin{trans} side of the pore, and the ones responsible for the anion selectivity, at the constriction part of the stem. We explained these effects, using ionic density maps, comparing the anions and cations behavior in the vicinity of the pore surface.

\section{Results and discussion}

All our CG-MD simulations are performed following the same protocol (see Methods) as in our previous study \cite{Basdevant2019}: we use the MARTINI CG force field for the $\alpha$HL protein inserted into a DPPC bilayer, surrounded by solvent (PW polarizable CG water molecules) and NaCl ions at a 1~M concentration. In addition, ElNeDyn elastic network~\cite{Periole2009} was applied on the protein structure.
We present the simulations of several systems for which we neutralized some of the charged amino acids on the seven chains of the $\alpha$HL protein. We refer to these simulations using the following numbering: e.g. D127N refers to the simulations for which the aspartic acid 127 of each chain was neutralized. 
Moreover, for each system, in order to mimic the voltage difference applied experimentally, several electric fields were applied perpendicularly to the membrane plan, ranging from $-0.04$~V/nm to $+0.04$~V/nm. It was shown that it is equivalent to applying a potential difference $\Delta V = E_z. L_z$~\cite{Gumbart2012}. Several previous MD studies on ionic transport through nanopores confirmed this relation~\cite{Modi2012, Aksimentiev2005, Aksimentiev2010, Crozier2001}.

From all these simulations, we generate the $IV$ curves of each pore and we calculated their ionic conductances $G^+$ and $G^-$ using two linear regressions, one for positive voltages and one for negative voltages. With the same method, we compute the separate conductances: $G_{\text{Cl}}$ for anions and $G_{\text{Na}}$ for cations.
For all our CG-MD simulations, we evaluate the current asymmetry using the conductance rectification ratio $r$ between positive and negative potentials: $r=\frac{G^+}{G^-}$ (see Methods for details). We also determine the anion asymmetry ratio: $r_{\text{Cl}}=\frac{G_{\text{Cl}}^+}{G_{\text{Cl}}^-}$, in order to identify whether the rectification effect comes from the anionic or cationic current. The ionic selectivity is defined by the selectivity ratio $s$ between the conductances of cations and anions for a positive electric potential: $s=\frac{{G_{\text{Na}}}^+}{{G_{\text{Cl}}}^+}$. An $s$ ratio equal to one stands for a non selective pore, if it is greater than one, the pore is cation selective, and if it is lower than one, the pore is anion selective. Table~\ref{TableSimPore} compiles rectification and selectivity ratios for all simulations, as well as the average conductances for positive and negative electric fields. We also draw the average ionic density maps for all simulations, in order to understand how charge distribution around the pore affects the ionic flow. This map is derived from the charged particle density averaged over the 1 to 1.5~$\mu$s part of the simulations, centered on the constriction part (see Methods).

\subsection{Wild-type and neutralized $\alpha$-hemolysins}

We first compare our results for the native heptameric protein (WT, wild-type) to a system with a totally neutralized pore, in order to validate our protocol. The WT pore carries a global charge of $+7$e, corresponding to the state of the protein in physiological conditions, whereas for the totally neutral pore, all protein partial charges were set to zero.

Without any restraint on the neutralized $\alpha$HL structure, we find almost no current passing through the pore. We observe a closure of the channel at the \latin{trans} extremity, particularly for low electric fields. We explain this effect by the lack of elastic network in this region. Since the elastic network is applied within a cut-off of 0.9~nm, the presence of loops at the \latin{trans} part of the channel is responsible for a gap between the different chains, which does not allow the addition of many springs. Therefore, this part of the structure is less stabilized by the elastic network.
For the native protein, coulombic repulsion between charged amino acids helps to maintain the channel open. On the contrary, for the neutral pore, the \latin{trans} extremity is unstable and collapses. 
We observe the same closure of the \latin{trans} extremity in other CG-MD simulations with a modified $\alpha$HL at the \latin{trans} side (especially D127N and D127N-D128N-K131N~; data not shown). To avoid this closure effect, for all systems with a pore modified at the \latin{trans} extremity, position restraints are applied on the protein CG backbone beads to their initial positions. Pore simulations using these position restraints are called thereafter "constrained pores", or distinguished by an asterisk (*).

\begin{figure}
\centering
\topinset{\bfseries \large{\helvet{(A)}}}{\includegraphics[width=0.49\textwidth]{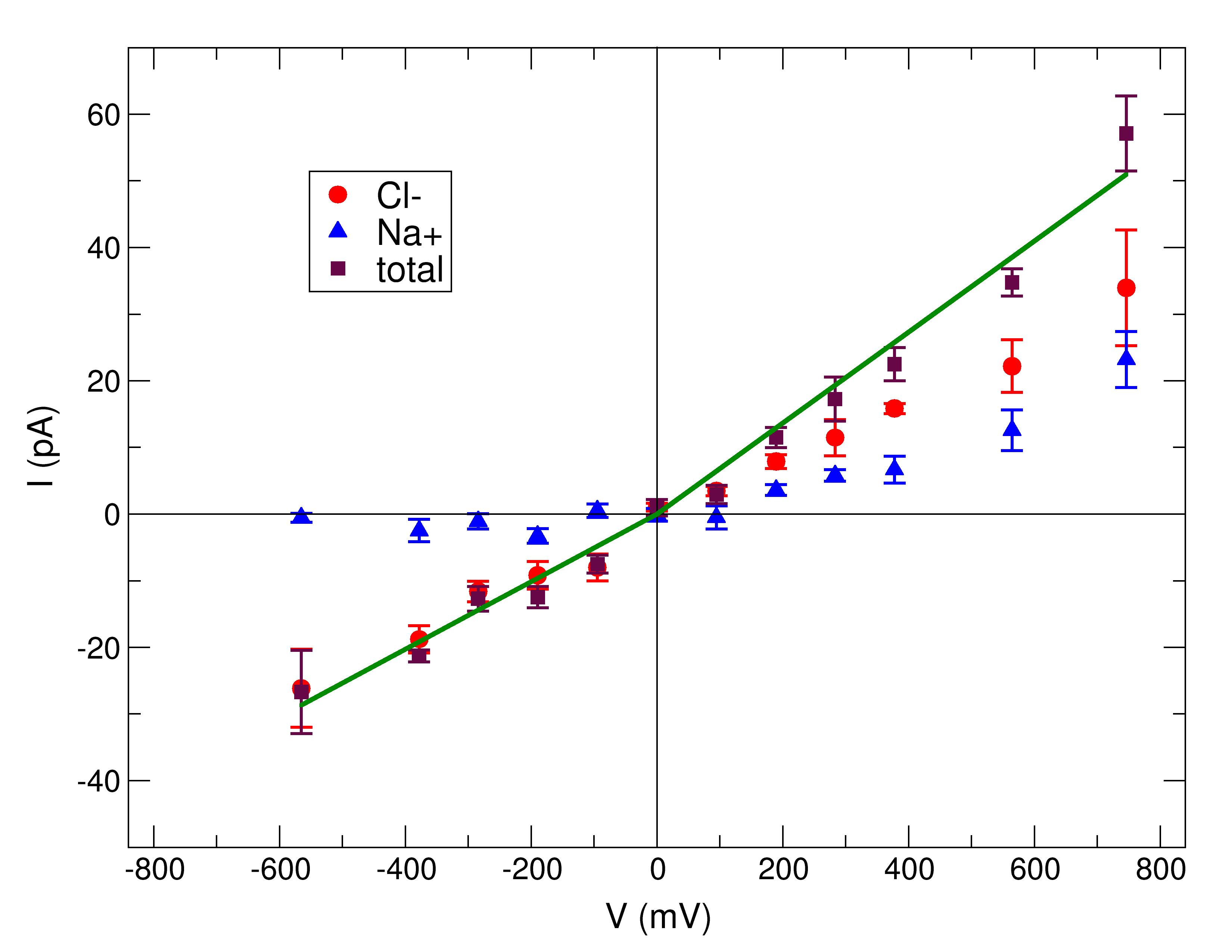}}{0cm}{-3.9cm}
\topinset{\bfseries \large{\helvet{(B)}}}{\includegraphics[width=0.49\textwidth]{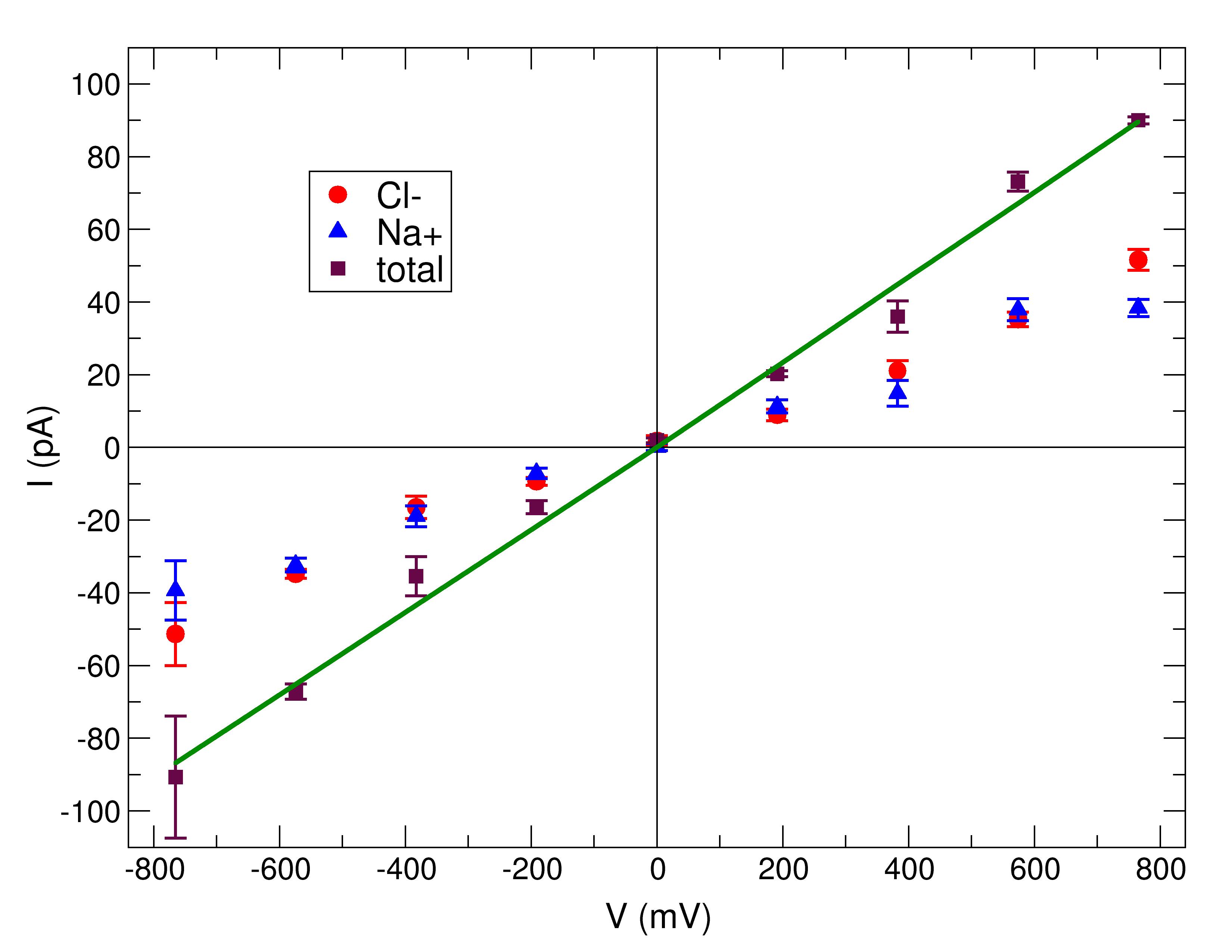}}{0cm}{-3.9cm}
\caption{$IV$ curves for the WT pore (A) and neutral constrained $\alpha$HL (B). We show in green the lines resulting from linear regression either on the positive or the negative potentials.}
\label{fig:iv-ch-ze}
\end{figure}

The current asymmetry for the WT pore was previously presented and briefly discussed~\cite{Basdevant2019}. It should be noted that, as we have also shown in our last paper, the CG currents are almost 15 times lower compared to experiments, which is due to the CG-particles size. 
Using the computed $IV$ curve of the WT pore (Figure \ref{fig:iv-ch-ze}A), we deduce a current asymmetry ratio of $r=1.35$ for the native pore at 1~M NaCl concentration, 320 K temperature and pH 7 (see Table~\ref{TableSimPore}). This value can be interpreted as a 35~$\%$ asymmetry, which is consistent with previous AA-MD studies~\cite{Aksimentiev2005,Bonome2017} and PNP and Brownian dynamics simulations~\cite{Noskov2004}. Experiments under the same conditions also show a 30\% asymmetry~\cite{Payet2015}. 
In Figure \ref{fig:iv-ch-ze}A, we see that the ionic current through the pore is mostly due to the chloride anions. However, the chloride conductances for positive and negative potentials are almost identical, unlike for the sodium cations, as confirmed by the chloride rectification factor $r_{\text{Cl}}$ in Table~\ref{TableSimPore} ($r_{\text{Cl}}=0.91$). Indeed, as shown on the $IV$ curve, the cation conductance for negative electric fields (${G^-_{\text{Na}}}$) is almost zero. Thereby, the current rectification effect measured for $\alpha$HL comes from the cationic current through the pore. 
Moreover, the selectivity ratio calculated from anion and cation conductances is $s=0.60$, showing that the WT $\alpha$HL is highly anion selective, as it was previously observed by both experiments and simulations~\cite{Menestrina1986, Walker1992, Gu2000, Noskov2004, Merzlyak2005, Mohammad2010, Simakov2010, Bhattacharya2011, Dyrka2013, Piguet2014, Boukhet2016, Bonome2017}.

We also studied the ionic current behavior for a neutralized constrained pore. Figure~\ref{fig:iv-ch-ze}B represents the corresponding $IV$ curve. We observe the absence of current rectification, highlighted by its current asymmetry ratio $r=1.03$, and a low anion selectivity ($s=0.85$). Indeed, contrary to the WT pore, a cationic current is measured for negative electric fields. Furthermore, cationic and anionic currents are almost equal, regardless of the sign of the electric field.
We can also notice that global current intensities are higher compared to the results for the WT pore with identical applied voltage (see conductance values in Table~\ref{TableSimPore}), which is confirmed by a scaling factor from experimental currents of only 8 instead of 15 for the WT pore. This is explained by a greater cationic current for the neutralized pore. These results are in accordance with the previous theoretical study of Noskov \latin{et al.} who obtained a current increase for systems with an $\alpha$-hemolysin neutralized at the top of the stem region~\cite{Noskov2004}.

\begin{figure}
\centering
\topinset{\bfseries \large{\helvet{(A)}}}{\includegraphics[trim={0.5cm 1cm 1cm 0},clip,scale=0.32]{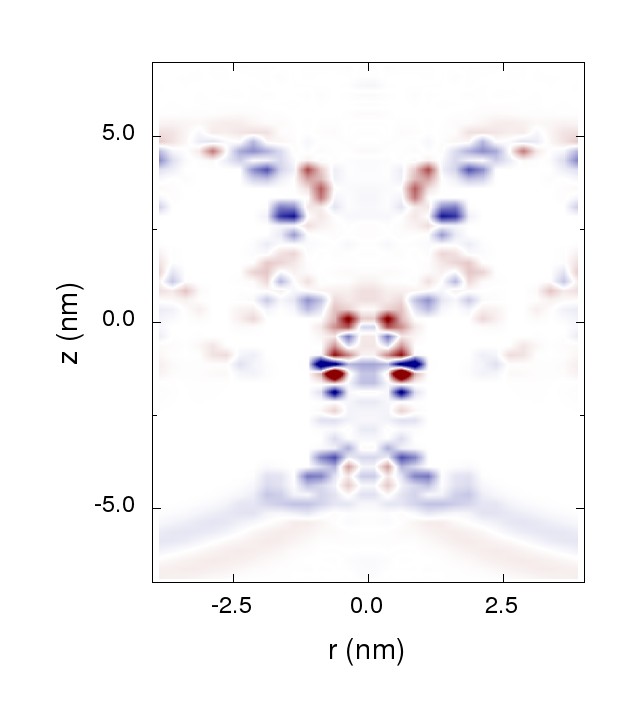}}{0.2cm}{-2.0cm}
\topinset{\bfseries \large{\helvet{(B)}}}{\includegraphics[trim={0.5cm 1cm 1cm 0},clip,scale=0.32]{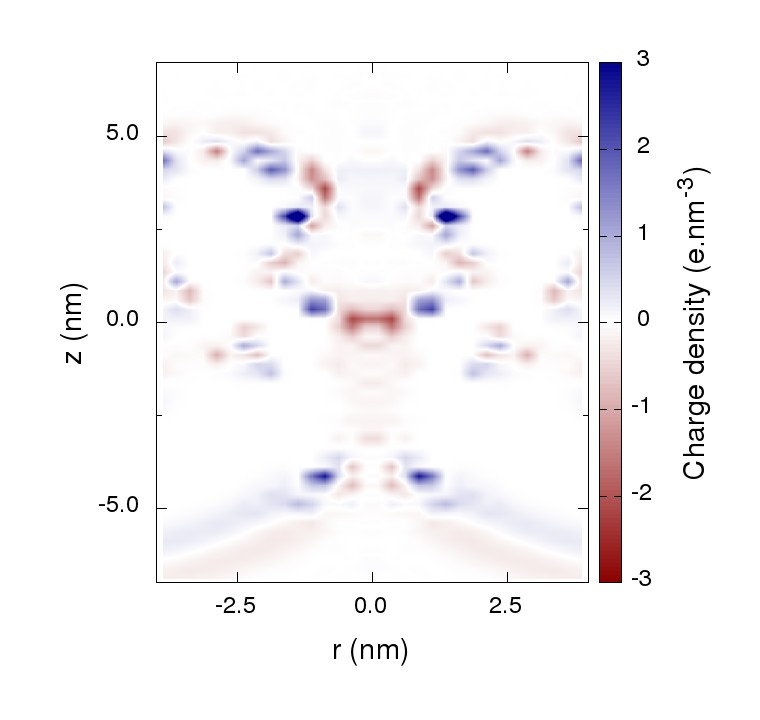}}{0.2cm}{-2.5cm}
\caption{Ionic density maps for the WT $\alpha$-hemolysin, in the presence of a positive electric field $E_z=+0.03$~V/nm (A) and negative electric field $E_z=-0.03$~V/nm (B).}
\label{fig:densmap_native}
\end{figure}

In addition, we plotted the average ionic density map for the WT pore, in order to locate the regions where the protein pore interacts strongly with ions. The average density is calculated on the last 500~ns of each trajectory (see Methods). Figure~\ref{fig:densmap_native} shows the average density maps for positive (A) and negative (B) electric fields ($E_z=\pm 0.03$~V/nm) of the WT pore. Ions accumulate at both extremities of the stem, for both positive and negative potentials. Sodium ions (in blue) are located at both sides, while chloride ions (in red) are mostly located just above the constriction part. Only a few ions are found inside the stem in the presence of a negative electric field. However, with a positive electric field, there are several interaction sites, mostly with sodium ions. This effect is consistent with the calculation of the total number of ions in the stem, showing many more ions when a positive electric field is applied (see Supplemental Information). Furthermore, these density maps point out that the difference between ionic densities with positive and negative electric fields mostly comes from the sodium ions. This result is in agreement with the fact that current rectification is due to the cationic current.

\subsection{$\alpha$-hemolysin with modified \latin{trans} side} 

It has previously been shown that the charged amino acids at the \latin{trans} side of $\alpha$HL are directly involved in the current rectification phenomenon: Noskov \latin{et al.}~\cite{Noskov2004} using 3D-PNP and Brownian Dynamics, Bhattacharya \latin{et al.}~\cite{Bhattacharya2011} using AA-MD simulations, Misakian and Kasianowicz performing pH experiments~\cite{Misakian2003} and Merzlyak \latin{et al.} \cite{Merzlyak2005} conducting mutagenesis experiments. We therefore computed the ionic current behavior of \emph{trans} amino acids modified pores.

We performed CG-MD simulations on seven $\alpha$HL with a modified \latin{trans} part and, as explained in the previous part of this article, position restraints are applied on the backbone of the protein for these simulations to prevent the pore closure. 

\begin{figure}
\centering
\topinset{\bfseries \large{\helvet{(A)}}}{\includegraphics[width=0.49\textwidth]{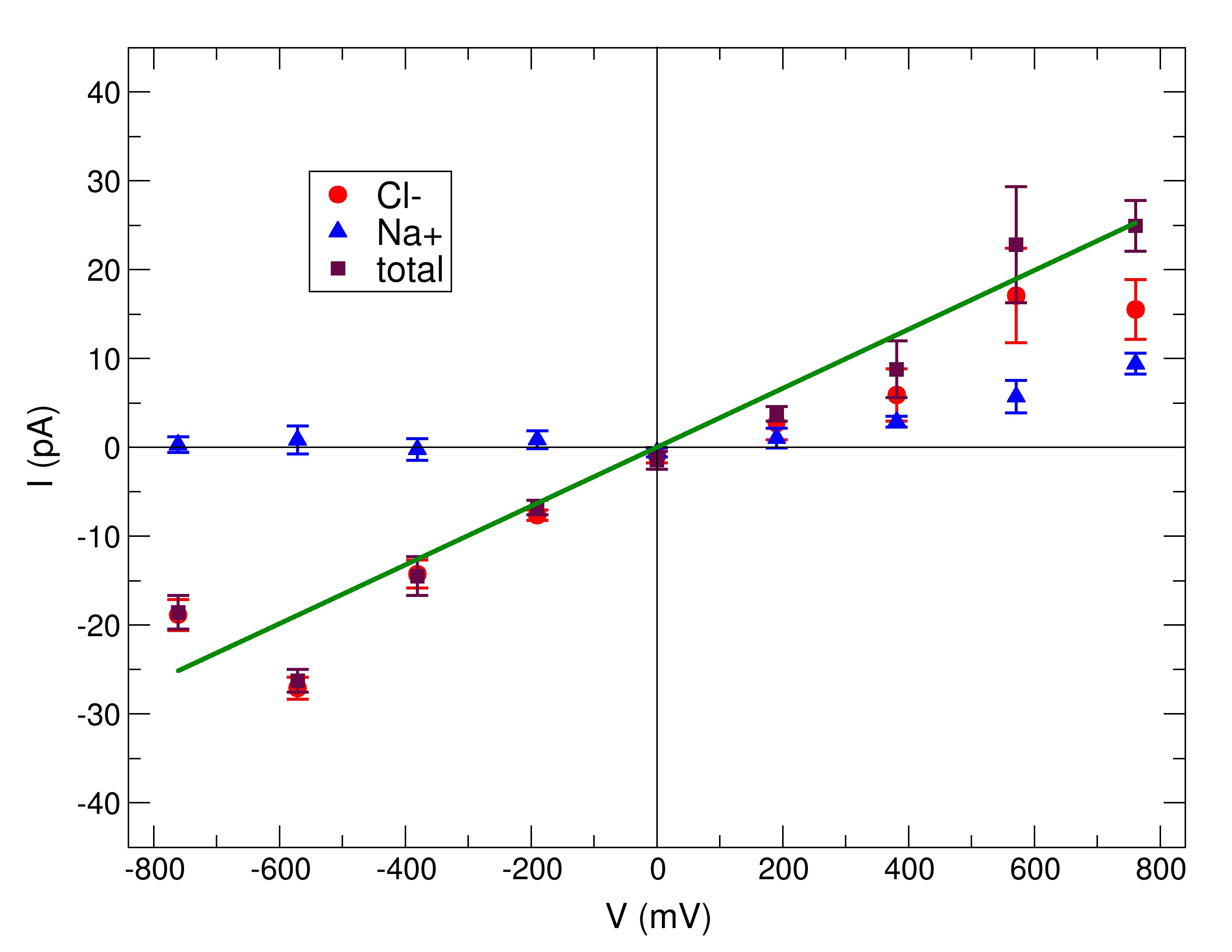}}{0cm}{-3.9cm}
\topinset{\bfseries \large{\helvet{(B)}}}{\includegraphics[width=0.49\textwidth]{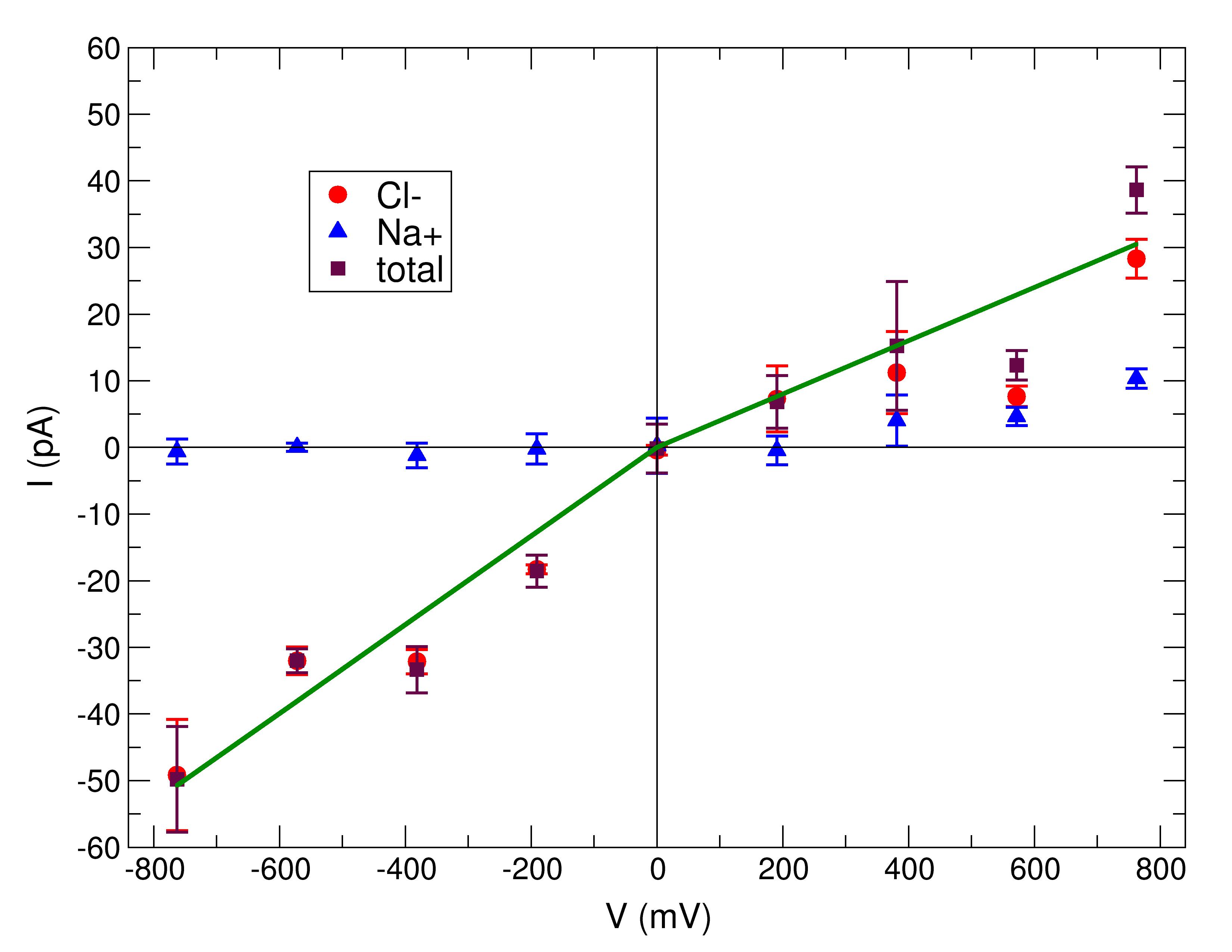}}{0cm}{-3.9cm}
\caption{$IV$ curves for the constrained $\alpha$HL with a modified \emph{trans} extremity: D128N* (A) and D127N-K131N* (B).}
\label{fig:iv-trans}
\end{figure}

The results for all our simulations with a modified \latin{trans} extremity are compiled in Table~\ref{TableSimPore}. 
We first focus on the charge influence of the aspartic acids D127 and D128 by neutralizing them separately, resulting in an $\alpha$HL with a neutral \latin{trans} part. We simulated thereby two systems: one with D127 protonated (D127N*) and one with D128 protonated (D128N*). 
Figure~\ref{fig:iv-trans}A shows the $IV$ curve obtained for D128N* system. No current rectification is observed for this pore ($r=1.0$), as well as a strong anion selectivity with a ratio $s=0.48$. Similarly to the WT pore, the cationic current is asymmetric and almost null for negative electric fields, but the anionic current shows a strong reverse asymmetry. The combination of these two currents results in no rectification. 
On the other hand, as shown on Table~\ref{TableSimPore}, for D127N*, the current asymmetry is reversed compared to the WT pore, with a ratio $r=0.60$, but the anion selectivity is still present ($s=0.40$). 
Our results highlight a different role for these two amino acids, and are consistent with the studies of Noskov \latin{et al.}~\cite{Noskov2004} and Bhattacharya \latin{et al.}~\cite{Bhattacharya2011} who found no rectification but a greater anion selectivity when the D128 were neutralized. 
Previous mutation experiments~\cite{Merzlyak2005} showed only a low decrease in rectification when D127 are mutated into cysteines, and a reverse asymmetry when D128 were mutated into cysteines. The fact that cysteines are smaller and that their polarity differs from neutral aspartic acids can explain the discrepancies between these experiments and our simulations using neutralized aspartic acids.
 
We also neutralized all \latin{trans} charged residues (D127, D128 and K131 on each of the seven monomers: D127N-D128N-K131N*). Table~\ref{TableSimPore} shows that there is almost no current rectification ($r=1.04$) for this pore, but a huge anion selectivity with a ratio $s=0.29$.
These results are consistent with previous theoretical studies~\cite{Noskov2004} and AA-MD~\cite{Bhattacharya2011}, and this tends to confirm the validity of our protocol. Our results also suggest that the rectification effect and the anion selectivity are not totally independent phenomena.

We simulated some additional \latin{trans} side modified pores: both D127 and D128 protonated (D127N-D128N*, with a \latin{trans} charge of $+7$e), only K131 deprotonated (K131N*, \latin{trans} charge of $-14$e), or K131N in combination with D127N or D128N, therefore with the same \latin{trans} charge than the WT-pore. 
We observe different behavior regarding the current rectification, while all these pores remained highly anion selective with ratios between $s=0.32$ and $s=0.49$, as presented in Table~\ref{TableSimPore}. Regarding rectification, as shown for D127N-K131N* in Figure~\ref{fig:iv-trans}B and D128N-K131N* in Table~\ref{TableSimPore}, the absence of charge on K131 in combination with the neutralization of one of the two aspartic acids results in a reverse asymmetry (respectively $r=0.60$ and $r=0.61$), although the total \latin{trans} charge is the same as the WT $\alpha$HL. For these two systems, this effect comes from a greater anionic current for negative electric fields compared to WT. 
For D127N-D128N* and for K131N*, a loss of current asymmetry was obtained ($r=1.08$ and $r=0.96$, respectively). These results suggest that this is not the total charge of the \latin{trans} part of the stem which is responsible for rectification, but the delicate balance between the charges of the three charged amino acids which can influence the ionic distribution in the stem, as will be discussed thereafter.

\begin{figure}
\centering
\topinset{\bfseries \large{\helvet{(A)}}}{\includegraphics[trim={0.5cm 1cm 1cm 0},clip,scale=0.32]{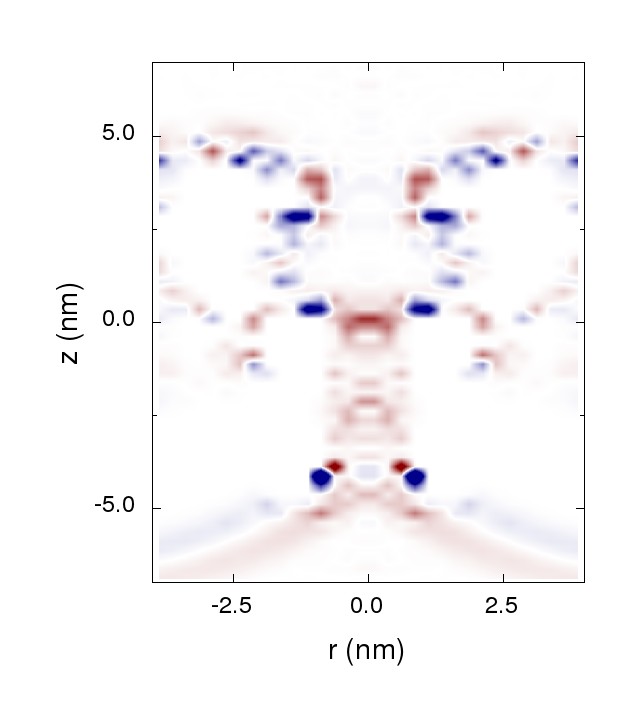}}{0.2cm}{-2.0cm}
\topinset{\bfseries \large{\helvet{(B)}}}{\includegraphics[trim={0.5cm 1cm 1cm 0},clip,scale=0.32]{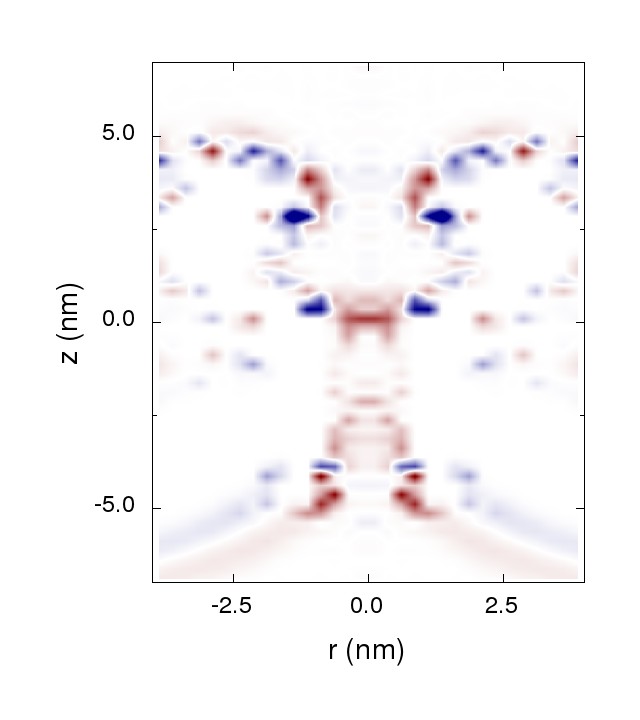}}{0.2cm}{-2.0cm}
\topinset{\bfseries \large{\helvet{(C)}}}{\includegraphics[trim={0.5cm 1cm 1cm 0},clip,scale=0.32]{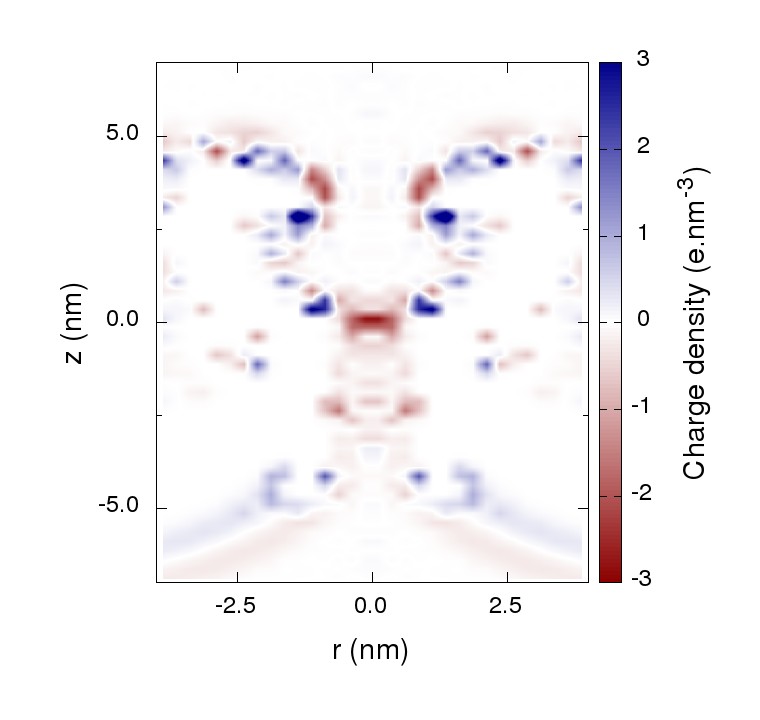}}{0.2cm}{-2.5cm}
\caption{Ionic density maps for the modified $\alpha$-hemolysins D128N* (A), D127N* (B) and D127N-K131N (C) in the presence of a negative electric field $E_z=-0.03$~V/nm.}
\label{fig:densmap_Asp128P}
\end{figure}

Figure~\ref{fig:densmap_Asp128P} represents average ionic density maps, in the presence of a negative electric field of $E_z=-0.03$~V/nm (differences in the density maps are more noticeable with negative fields), for the D128N* (A), D127N* (B) and D127N-K131N* (C) $\alpha$HL pores. 
D128N* shows no rectification, while D127N* and D127N-K131N* present an inverse rectification.
Compared to the WT pore, we observe a higher negative ionic density inside the stem for these three systems in the presence of a negative electric field. This is consistent with a null cationic current at negative voltages for all this pores compared to the already weak cationic current of the WT.

All these density maps are very similar at the constriction part. However, D128N* (Fig.~\ref{fig:densmap_Asp128P}A) and D127N* (Fig.~\ref{fig:densmap_Asp128P}B) display very different characteristics of ionic densities at the \latin{trans} part, despite sharing the same charge. Indeed, D127 residues point towards the inner pore and are very close to K131, while the D128 amino acids point towards the outer pore (see Fig.~\ref{fig:aHL_charges}). D127 and K131 tends to form salt bridges, and the neutralization of one of these residues breaks this bond, therefore affecting the current asymmetry and the ionic density. 

For the D128N* pore, we notice on Fig.~\ref{fig:densmap_Asp128P}A two peaks of positive and negative ionic densities of similar intensity at the \latin{trans} part. Since the total charge for this part of the protein is zero, the resulting global charge of the system at the \latin{trans} part is almost null. This is consistent with the lack of current rectification of this pore. 

For the D127N* system, we observe on Fig.~\ref{fig:densmap_Asp128P}B a highly negative spot at the \latin{trans} side of the stem, which is not present for the other pores, including the WT. The accumulation of anions at this location compensates the positive charges of the K131 residues. The bottom global charge is therefore negative.

The D127N-K131N* pore on Fig.~\ref{fig:densmap_Asp128P}C, bearing the same charge as the WT $\alpha$HL, does not reveal the positive density peak at the \latin{trans} part observed for WT pore (see Figure~\ref{fig:densmap_native}). Thus, the -7$e$ negative charge at the bottom of the pore is not compensated by the presence of cations, resulting in a bottom global charge similar to D127N*. The presence of this negative global charge at the \latin{trans} part causes the reverse rectification phenomenon. 
We should notice that, whereas D127N* and D127N-K131N* systems both exhibit the same reverse rectification, the currents are twice as low for D127N*. We can explain this phenomenon by the absence of ion accumulation at this side of the D127N-K131N* pore which induces a higher conductance compared to D127N*.

The density maps of the other $\alpha$HL pores can be found in the Supplemental Information. The study of these ionic distributions confirms the role of the \latin{trans} part charges on current rectification, and provides an explanation for the different rectification behaviors of the pores despite similar total charges.

Our analyses confirm that the three charged amino acids of the $\alpha$HL \latin{trans} are responsible for the rectification but play no direct role in the anion selectivity. We also identified that D127 and D128 play different roles because of their different positioning. As neutralizing some amino acids at the \latin{trans} part (D128N*, D127N-D128N-K131N*, K131N* and D127N-D128N*) cancels the rectification effect, the neutralization of certain combinations of amino acids can reverse the current asymmetry (D127N*, D127N-K131N* and D128N-K131N*). These behaviors can be explained by the distribution of ions within the channel, apparent in our ionic density maps. All these systems keep an anion selectivity equivalent or greater than the WT pore. 

\subsection{$\alpha$-hemolysin with modified constriction} 

In this section, we focus on the charged residues located in the $\alpha$HL constriction, at the top of stem: E111 and K147 (see Fig.~\ref{fig:aHL_charges}). The constriction is the narrower part of the pore, where the channel diameter reduces from 4.6 to 1.4~nm. It carries a null total charge for the WT pore and, since the two charged amino acids form a salt bridge, a modification of the charges of these residues should influence the ionic flow. 
A theoretical study from Noskov and Roux~\cite{Noskov2004} showed that the charges at the constriction influence anion selectivity and rectification. Moreover, several experimental mutagenesis studies revealed the role of these amino acids in anion selectivity: Gu \latin{et al.}~\cite{Gu2000} measured a weak cation selectivity when E111 and K147 were mutated into neutral asparagines; Maglia \latin{et al.}~\cite{Maglia2008} found that the mutation of E111 into asparagines yields a slight increase into anion selectivity; Mohammad and Movileanu~\cite{Mohammad2010} observed a cation selectivity when K147 were mutated into negatively charged aspartic acids. 

We constructed three systems with modified charges: with protonated glutamic acids E111 (E111N, resulting in a total top charge of $+7e$) with deprotonated lysines K147 (K147N, top charge $-7e$), and with both amino acids neutralized (E111N-K147N, null top charge as the WT-pore).

\begin{figure}
\centering
\hfill
\topinset{\bfseries \large{\helvet{(A)}}}{\includegraphics[width=0.47\textwidth]{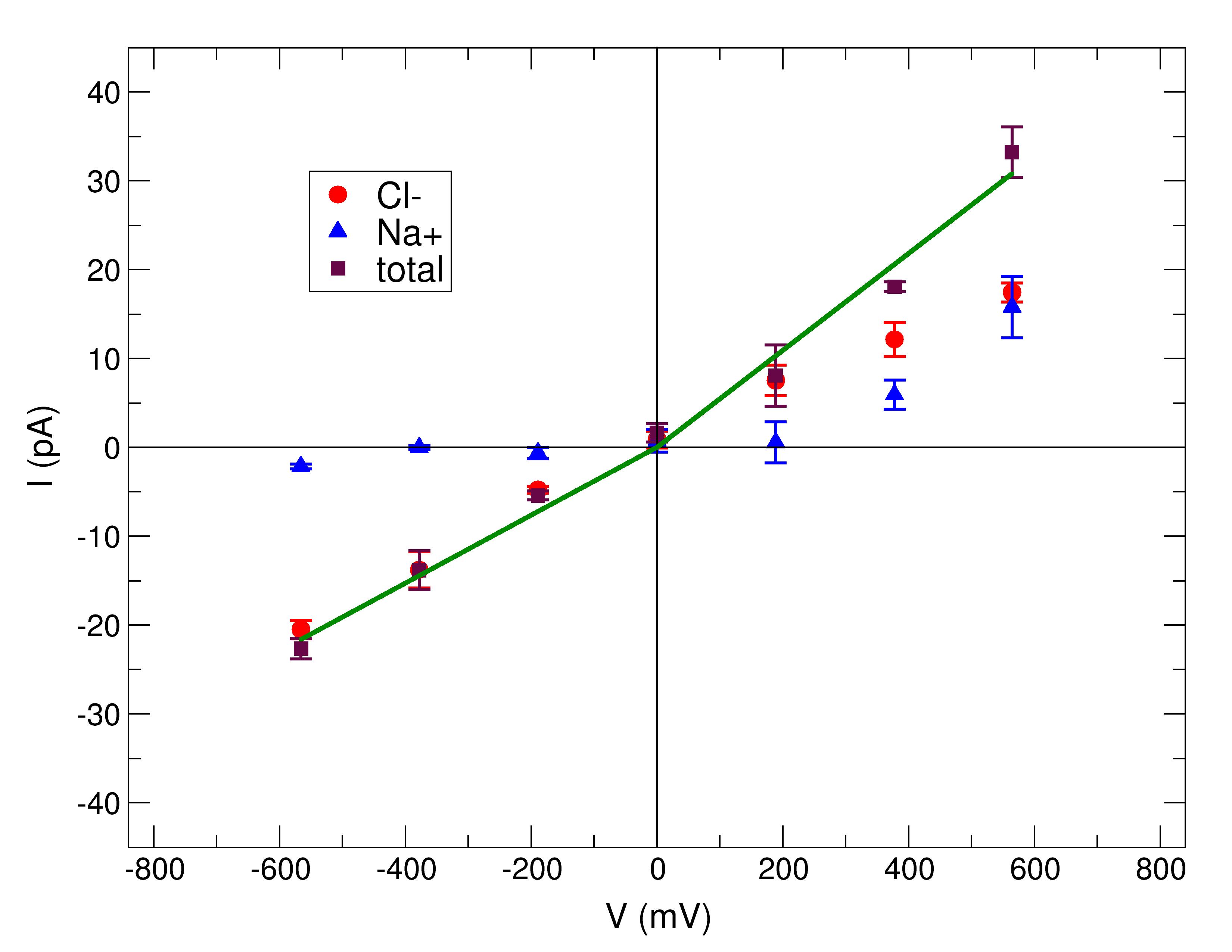}}{0cm}{-3.9cm}
\topinset{\bfseries \large{\helvet{(B)}}}{\includegraphics[width=0.47\textwidth]{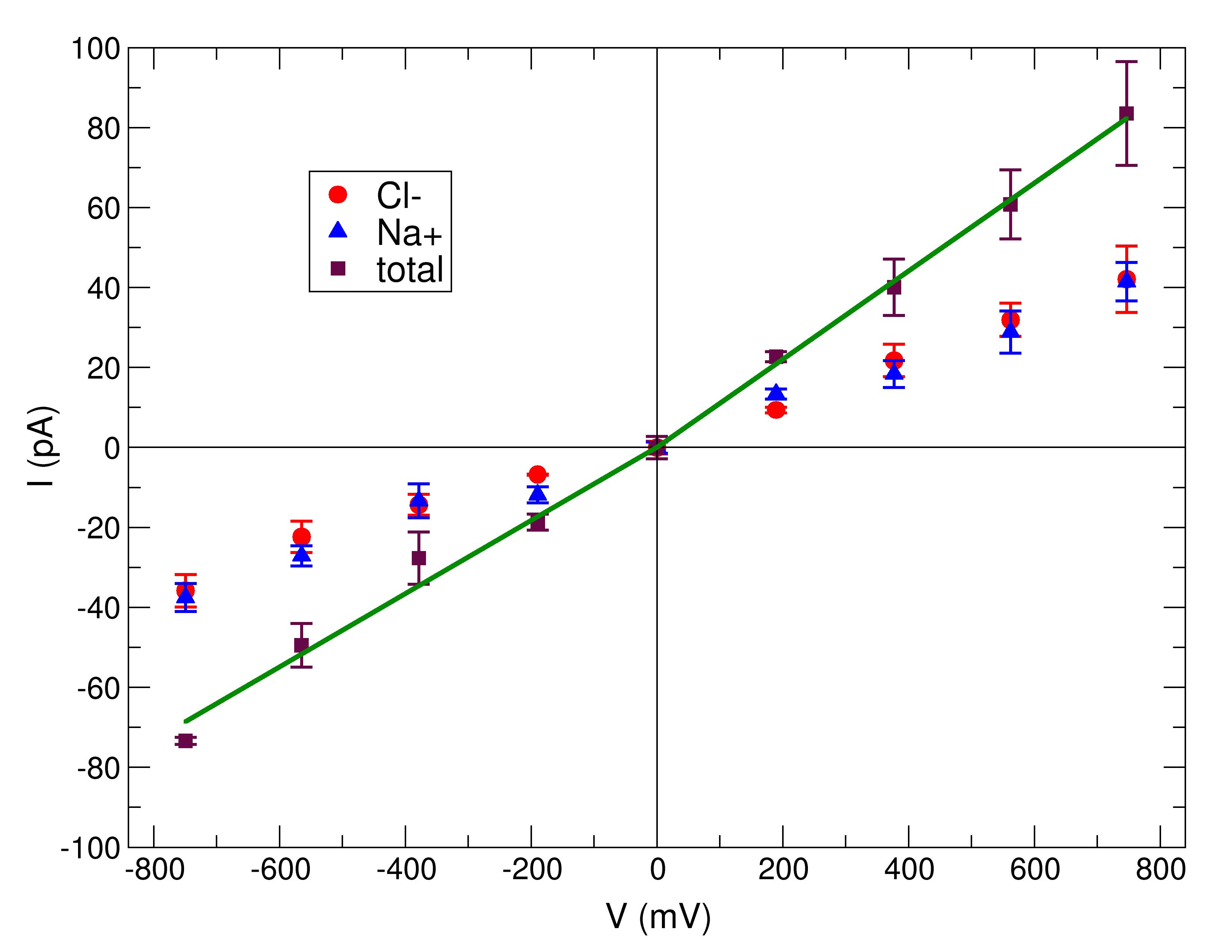}}{0cm}{-3.9cm}
\caption{$IV$ curves for $\alpha$HL with a modified constriction: E111N (A) and K147N (B).}
\label{fig:iv-topstem}
\end{figure}

\begin{figure}
\centering
\topinset{\bfseries \large{\helvet{(A)}}}{\includegraphics[trim={0.5cm 1cm 1cm 0},clip,scale=0.32]{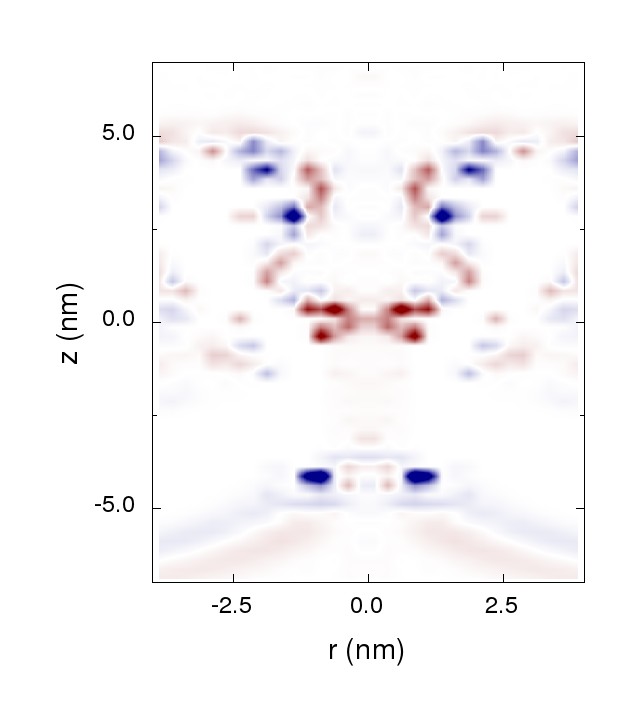}}{0.2cm}{-2.0cm}
\topinset{\bfseries \large{\helvet{(B)}}}{\includegraphics[trim={0.5cm 1cm 1cm 0},clip,scale=0.32]{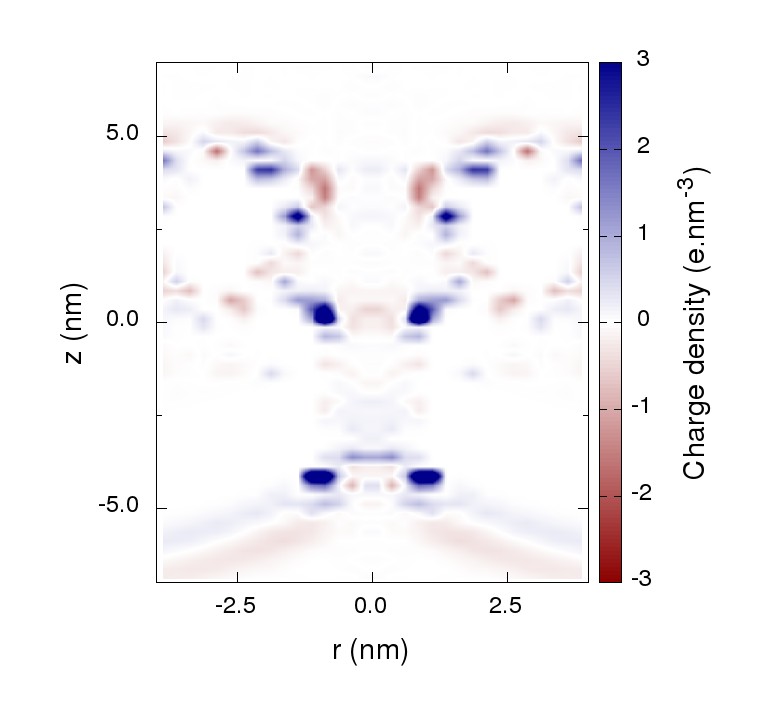}}{0.2cm}{-2.5cm}
\caption{Ionic density maps for the modified $\alpha$-hemolysins E111N (A) and K147N (B) in the presence of a negative electric field $E_z=-0.03$~V/nm.}
\label{fig:densmap_Lys147dP}
\end{figure}

Figure~\ref{fig:iv-topstem} shows the $IV$ curves obtained for the E111N (A) and K147N (B) systems. 
As observed on Fig.~\ref{fig:iv-topstem}A and Table~\ref{TableSimPore}, E111N $\alpha$HL presents a current asymmetry with a ratio $r=1.43$ and an anion selectivity similar to the native protein ($s=0.71$), therefore suggesting that E111 has almost no effect on selectivity or rectification. Yet, it should be noticed that, for this system, the anchoring of the pore inside the membrane is weakened, resulting in system crashes for high electric fields ($E_z=\pm 0.04$~V.nm$^{-1}$). This result points out the important role of E111 in the stability of the pore insertion inside the membrane. 
The total current for E111N is smaller than the WT-protein, as shown by conductances in Table~\ref{TableSimPore}. This result is not consistent with previous experiments of E111 mutations into asparagines for which the total current slightly increased~\cite{Maglia2008}. However, this discrepancy can be explained, as suggested by Stoddard \latin{et al.}~\cite{Stoddart2010}, by the fact that asparagines are smaller residues than glutamic acids and the mutated pore diameter is therefore a little wider than the WT at the constriction. The difference in polarity between these residues can also explain this difference. Asparagines, due to their polarity, do probably not allow the accumulation of anions at the constriction as for the neutral glutamic acids (as explained below and visible on Fig.~\ref{fig:densmap_Lys147dP}A), resulting in a larger diameter and a higher current.

Figure~\ref{fig:iv-topstem}B shows that the K147N pore presents a slightly weaker current rectification than the WT-pore ($r=1.21$) and no ionic selectivity ($s=0.96)$. Cationic and anionic currents are both asymmetric in this case ($r_{\text{Cl}}=1.29$), contrary to the WT pore for which the Cl$^-$ current is nearly symmetric. Therefore, there is a rectification effect even without selectivity.
This result suggests that K147 plays a determinant role in the ionic selectivity of $\alpha$HL. This is in accordance with a previous mutagenesis study~\cite{Mohammad2010} which mutated K147 into aspartic acids and measured a cation selectivity.

Finally, the E111N-K147N pore looses the ionic selectivity but also the current asymmetry ($s=1.01$, $r=0.99$). The absence of rectification in this case is due to both symmetric cation and anionic currents, whereas for the WT-pore, rectification is due to the asymmetry in the Na$^{+}$ current. It is in accordance with a previous experimental study where E111 and K147 were mutated to asparagines and current rectification disappeared~\cite{Rincon-Restrepo2011}. We can again notice that anion selectivity and rectification effects are intertwined, and the modification of one feature can influence the other.

The total current is greater for the two systems in which K147 was deprotonated compared to the WT-pore, showing conductances close to the neutralized pore (see Table~\ref{TableSimPore}). Since these systems present no anion selectivity, the cationic current is greater than for the WT pore, yielding a higher total current. This is consistent with several previous theoretical~\cite{Noskov2004} and experimental studies~\cite{Rincon-Restrepo2011, Ayub2012}, but other experiments~\cite{Gu2000, Stoddart2009, Stoddart2010} did not measure any increase of the ionic flow.

The average ionic density maps of the E111N (A) and K147N (B) systems, with a negative electric field, are plotted on Figure~\ref{fig:densmap_Lys147dP}. 
Both systems present a current rectification like the WT-pore, but K147N show no anion selectivity, whereas E111N is anion selective. Indeed, we observe a completely different picture at the constriction between the two pores. 
The E111N (Fig.~\ref{fig:densmap_Lys147dP}A) shows a very high anion density around the constriction. This anion distribution compensates the global positive charge of this region and prevents cations from passing, hence responsible for the anion selectivity. 
On the contrary, for the K147N, as shown on Fig.~\ref{fig:densmap_Lys147dP}B, there are no anions at the constriction site, and a high cation density is observed. The cations are allowed to pass in this case, resulting in the absence of anion selectivity. In addition, the channel is wider for K147N than for E111N, explaining that the total current is greater than WT and E111N. 

At the \latin{trans} part, with a negative potential, the two maps reveal a positive peak similar to the WT-pore. There are more cations located at the \latin{trans} extremity, which can explain that the rectification effect is still observed for K147N, although we could expect that the presence of a cationic current with negative electric fields would induce the absence of rectification (the anionic current is much smaller for negative electric fields, as shown by the $r_\text{Cl}=1.29$ ratio).
Consequently, we remarked that charge modifications at one stem extremity can influence the ionic density not only close to the charges but also at the opposite extremity. For all non selective pores, no anions were found at the constriction. Furthermore, with negative electric fields, there is an increase in the number of cations at the \latin{trans} side (see Supplemental Information for the total number of ions in the stem of K147N).

Our results confirm the role of the K147 charges in the anion selectivity of $\alpha$HL. We also found that, like with modified charges at the \latin{trans} part, and according to the ionic density maps, it is the detailed residue charges of the constriction part which are responsible for the anion selectivity, and not the global charge of the restriction region. \\

\begin{table}[htb]
\centering
\caption{Conductance values from CG-MD simulations calculated from linear regression on $IV$ curves. The total charge of the pore, the asymmetry ratio $r=\frac{G^+}{G^-}$, the selectivity ratio $s=\frac{G_{\text{Na}}^+}{G_{\text{Cl}}^+}$ and the asymmetry ratio for the anionic current $r_{\text{Cl}}$ are indicated for each protein. The selectivity ratio $s$ was evaluated using conductances for simulations with a positive voltage (+).}

\begin{tabular}{|c|c|c|c|c|c|c|}
\hline
 & $Q_{\text{tot}}$ & $G^+$  & $G^-$  & $r$  & $s$ & $r_{\text{Cl}}$ \\
  & (e) &  (pS) &  (pS) & $G^+/G^-$ &  $G_{\text{Na}}^+/G_{\text{Cl}}^+$ & $G_{\text{Cl}}^+/G_{\text{Cl}}^-$ \\
 \hline
 \textbf{WT}                 & +7   & 68.3 & 50.7 & 1.35 & 0.60 & 0.91 \\
 \textbf{Neutral*}           & 0    & 117.1 & 113.5 & 1.03 & 0.85 & 1.03 \\
 \hline
 \textbf{D127N*}             & +14  & 20.3 & 33.8 & 0.60 & 0.40 & 0.43 \\
 \textbf{D128N*}             & +14  & 33.2 & 33.1 & 1.00 & 0.48 & 0.66 \\
 \textbf{D127N-D128N-K131N*} & +14  & 36.2 & 34.7 & 1.04 & 0.29 & 0.88 \\
 \textbf{K131N*}             & 0    & 33.8 & 35.2 & 0.96 & 0.49 & 0.65 \\
\textbf{D127N-K131N*}        & +7   & 40.0 & 66.5 & 0.60 & 0.38 & 0.44 \\
\textbf{D128N-K131N*}        & +7   & 33.9 & 55.2 & 0.61 & 0.32 & 0.48 \\
\textbf{D127N-D128N*}        & +21  & 38.0 & 35.1 & 1.08 & 0.40 & 0.80 \\
\hline
 \textbf{E111N-K147N}        & +7   & 120.0 & 119.4 & 1.00 & 0.99 & 1.01 \\
 \textbf{E111N}              & +14  & 54.6 & 38.2 & 1.43 & 0.71 & 0.90 \\
 \textbf{K147N}              & 0    & 110.3 & 91.5 & 1.21 & 0.96 & 1.29 \\

\hline

\end{tabular}

* Simulations run with Position Restraints on the backbone beads

\label{TableSimPore}
\end{table}

\section{Conclusion}

In this study, we characterized the role of certain amino acids of the $\alpha$-hemolysin pore using coarse-grained molecular dynamics. By computing $IV$-curves and ionic density maps of a very exhaustive panel of $\alpha$HL charge modifications, we could reproduce experimentally observed behaviors as well as all-atom MD simulations results. We demonstrated that the current rectification is due to the \latin{trans} side charges, in particular D128. On the other hand, we showed that the anion selectivity arises from the K147 residues located at the \latin{cis} constriction. Even though these two effects are closely linked, they are essentially due to subtle changes of the electrostatic potential and not to the global charge of the pore or of the channel extremities. The use of ionic density maps gives some microscopic insight of these phenomenon confirming the importance of the local potential due to D128 and K147 in, respectively, current rectification and anion selectivity.

This study demonstrates the ability of CG-MD, with a polarizable solvent, to reliably model nanopore currents using voltages closer to experimental conditions and smaller than AA simulations. It can describe the molecular insight of processes with a timescale getting closer to translocation experiments and it allows the study of numerous mutations. This modeling system could be used to model nanopore molecular transport such as DNA translocation or could be applied to other nanopores like aerolysin.

\section{Methods}

\subsection{Simulations setup}

Different systems were studied, either with a native pore or with a modified $\alpha$-hemolysin. They were all prepared with the same protocol whose first steps are identical to our previous paper~\cite{Basdevant2019}.

First, the crystallographic structure of $\alpha$-hemolysin was taken from the Protein Data Bank (PDB entry: 7AHL)~\cite{Berman2000} The coordinates of the missing atoms were recovered using the \emph{pdb2pdr} software~\cite{Dolinsky2004} and the water molecules were removed. Using the \emph{propka} software~\cite{Olsson2011}, we determined the protonation states of charged residues: all histidines were therefore considered neutral. 

The protein was reduced in coarse grains using the \emph{martinize.py} script provided by MARTINI, with the Martini 2.2 force field~\cite{deJong2013}. The ElNeDyn approach from MARTINI was applied on the protein~\cite{Periole2009}, adding an elastic network on the backbone and between the different chains of the $\alpha$-hemolysin, which consists of springs with 500~kJ.mol$^{-1}$.nm$^{-2}$ elastic bond strength connecting grains within a 0.9~nm cutoff from each other.

After a short minimization, the backbone of the resulting protein structure was fitted on the one of a coarse-grained $\alpha$-hemolysin inserted in a dipalmitoylphosphatidylcholine (DPPC) bilayer obtained by Pr. Sansom~\cite{Scott2008}. This last protein structure was then deleted resulting in a system composed of our $\alpha$-hemolysin structure inserted in a DPPC bilayer of 756 molecules in a simulation box initially set at 15.85 x 15.85 x 20~nm. 
A short minimization was performed and the system was solvated with 28,758 W MARTINI CG water molecules using the \emph{genbox} option from the Gromacs package (version 4.6). 
Another minimization was then carried out followed by two equilibration steps: one of 1~ns in the NVT ensemble (using a 10~fs timestep) and the other of 5~ns in the NPT ensemble (using a 20~fs timestep).\\

Afterwards, the W MARTINI water was transformed into polarizable water PW~\cite{Yesylevskyy2010} using the \emph{triple\_w.py} script provided by MARTINI, and the system underwent a short minimization.
The charges of the amino acids of the protein were then modified as required and we performed two equilibration steps (using Gromacs package version 5.1.4): 2~ns at NVT and 10~ns at NPT, both with position restraints on the protein backbone. Next, we added the ions with the \emph{genion} tool of Gromacs in order to obtain a neutral system with 1~M ionic concentration. For example, for a neutral $\alpha$-hemolysin, we added 2054 of both Na$^+$ and Cl$^-$ MARTINI ions. We carried out another equilibration step of 10~ns at NPT with position restraints on the protein backbone. The resulting systems were used as initial state for the simulations. \\

It should be noted that the MARTINI ions named Na$^+$ and Cl$^-$ are totally symmetrical in terms of mass ($m=72$~g.mol$^{-1}$) and size, but carry opposite charges.
We can therefore compare our coarse-grained simulations to experiments or all-atom simulations with KCl ions.

\subsection{MD methods}
All our molecular dynamics simulations were performed using the GROMACS package programs (version 5.1.4) with the MARTINI force field (martini-v.2.2P). Periodic boundaries conditions were applied on the system. 
We used Particle-mesh Ewald (PME) method for long-range electrostatics calculations, with a 2~\AA\ Fourier grid spacing and a direct space cut-off radius of 13~\AA. Simulations were performed on the NPT ensemble, at a temperature of 320~K and 1~bar pressure. Nos\'e-Hoover thermostat and Parinello-Rahman barostat were applied. A timestep of 20~fs was used for all the simulations.

Different electric fields $E_z$ were applied in the $z$ direction, perpendicularly to the membrane plan, for each system, going from $-0.04$~V.nm$^{-1}$ to $+0.04$~V.nm$^{-1}$: 0, $\pm 0.01$, $\pm 0.02$, $\pm 0.03$ and $\pm 0.04$~V.nm$^{-1}$. Each simulation was conducted for 1.5~$\mu$s with a constant electric field applied. 

\subsection{Ionic current analyses}
For each applied electric field, we counted the ions crossing a horizontal plan at the \latin{trans} side of the pore which allowed us to obtain cumulative current over time, similarly to Refs~\cite{Guros2018,Basdevant2019}. The average current intensity was calculated from the slopes of the linear regressions on the cumulative current for 250~ns intervals with a shift of 50~ns. We computed this current intensity on 500~ns from 1 to 1.5~$\mu$s of simulation.
The applied voltage corresponding to each applied electric field was evaluated using $\Delta V={E_z}\times{L_z}$, with $\Delta V$ the voltage in V, $E_z$ the applied electric field in V.nm$^{-1}$ and $L_z$ the average length of the box in $z$ direction, during the relevant time window, in nm.
From these calculations of average current intensities and applied voltages, we plotted the characteristic $IV$ curves of the pores, corresponding to the current intensities as a function of the voltage applied on the system.

Intensities generated by either the Na$^+$ or the Cl$^-$ flows were also obtained as a function of the applied electric fields, by the same method, and were represented separately on the $IV$ curves.

The positive ($G^+$) and negative ($G^-$) conductances were computed using the slope of linear regressions on these $IV$ curves for the positive or negative voltages, respectively.

The current asymmetry was calculated using the ratio $r=\frac{G^+}{G^-}$ and the ionic selectivity was deduced from the ratio $s=\frac{{G_{Na^+}}^+}{{G_{Cl^-}}^+}$ with ${G_{Na^+}}^+$ and ${G_{Cl^-}}^+$ the positive conductances for each ion species.

\subsection{Ionic charge densities analyses}

We used \emph{densmap} tool from Gromacs package to plot the average ionic densities of the different systems from 1 to 1.5 µs of simulation. 
This tool can compute the average particle density during the simulation around an axis resulting in 2D-maps. Therefore, we defined the axis by using the residues at the bottom (D127, D128, T129, G130 and K131) and at the top (N17) of the pore. Moreover, we determined Na$^{+}$ and Cl$^{-}$ densities with a bin size of 0.25~nm and with the mirror option of \emph{densmap} to represent the density on both sides of the axis, leading to in a symmetric density centered on the pore axis. 

We calculated the charge density by combining Na+ and Cl- individual densities such as: $d = q_{Na} \times d_{Na} + q_{Cl} \times d_{Cl}$ with $q$ the ion charge and $d$ the density in one bin. Therefore, when there are more cations than anions in one bin, the charge density is positive and conversely for negative density.

\begin{acknowledgement}

 This work was granted access to the HPC resources of CINES under the allocations A0020707139 (2017), A0040707139 (2018) and A0060707139 (2019) made by GENCI.

\end{acknowledgement}

\begin{suppinfo}

Number of ions in function of applied voltage on WT and K147N $\alpha$HL pores. $IV$ curves for D127ND128N-K131N*, D127N*, K131N*,
D127N-D128N*, D128N-K131N* and E111N-K147N $\alpha$HL. Ionic density maps for the neutral constrained pore and for D127N-D128N-K131N*, E111N-K147N, K131N*, D127N-D128N* and D128N-K131N* $\alpha$HL.

\end{suppinfo}

\bibliography{CGNanopores}

\providecommand{\latin}[1]{#1}
\makeatletter
\providecommand{\doi}
  {\begingroup\let\do\@makeother\dospecials
  \catcode`\{=1 \catcode`\}=2 \doi@aux}
\providecommand{\doi@aux}[1]{\endgroup\texttt{#1}}
\makeatother
\providecommand*\mcitethebibliography{\thebibliography}
\csname @ifundefined\endcsname{endmcitethebibliography}
  {\let\endmcitethebibliography\endthebibliography}{}
\begin{mcitethebibliography}{54}
\providecommand*\natexlab[1]{#1}
\providecommand*\mciteSetBstSublistMode[1]{}
\providecommand*\mciteSetBstMaxWidthForm[2]{}
\providecommand*\mciteBstWouldAddEndPuncttrue
  {\def\EndOfBibitem{\unskip.}}
\providecommand*\mciteBstWouldAddEndPunctfalse
  {\let\EndOfBibitem\relax}
\providecommand*\mciteSetBstMidEndSepPunct[3]{}
\providecommand*\mciteSetBstSublistLabelBeginEnd[3]{}
\providecommand*\EndOfBibitem{}
\mciteSetBstSublistMode{f}
\mciteSetBstMaxWidthForm{subitem}{(\alph{mcitesubitemcount})}
\mciteSetBstSublistLabelBeginEnd
  {\mcitemaxwidthsubitemform\space}
  {\relax}
  {\relax}

\bibitem[Kasianowicz \latin{et~al.}(1996)Kasianowicz, Brandin, Branton, and
  Deamer]{Kasianowicz1996}
Kasianowicz,~J.~J.; Brandin,~E.; Branton,~D.; Deamer,~D.~W. Characterization of
  Individual Polynucleotide Molecules Using a Membrane Channel. \emph{Proc.
  Natl. Acad. Sci. U. S. A.} \textbf{1996}, \emph{93}, 13770--13773\relax
\mciteBstWouldAddEndPuncttrue
\mciteSetBstMidEndSepPunct{\mcitedefaultmidpunct}
{\mcitedefaultendpunct}{\mcitedefaultseppunct}\relax
\EndOfBibitem
\bibitem[Dudko \latin{et~al.}(2010)Dudko, Mathé, and Meller]{Dudko2010}
Dudko,~O.~K.; Mathé,~J.; Meller,~A. Nanopore Force Spectroscopy Tools for
  Analyzing Single Biomolecular Complexes. \emph{Methods Enzymol.}
  \textbf{2010}, \emph{475}, 565--589\relax
\mciteBstWouldAddEndPuncttrue
\mciteSetBstMidEndSepPunct{\mcitedefaultmidpunct}
{\mcitedefaultendpunct}{\mcitedefaultseppunct}\relax
\EndOfBibitem
\bibitem[Meller \latin{et~al.}(2000)Meller, Nivon, Brandin, Golovchenko, and
  Branton]{Meller2000}
Meller,~A.; Nivon,~L.; Brandin,~E.; Golovchenko,~J.; Branton,~D. Rapid Nanopore
  Discrimination Between Single Polynucleotide Molecules. \emph{Proc. Natl.
  Acad. Sci. U. S. A.} \textbf{2000}, \emph{97}, 1079--1084\relax
\mciteBstWouldAddEndPuncttrue
\mciteSetBstMidEndSepPunct{\mcitedefaultmidpunct}
{\mcitedefaultendpunct}{\mcitedefaultseppunct}\relax
\EndOfBibitem
\bibitem[Oukhaled \latin{et~al.}(2007)Oukhaled, Math\'e, Biance, Bacri, Betton,
  Lairez, Pelta, and Auvray]{Oukhaled2007}
Oukhaled,~G.; Math\'e,~J.; Biance,~A.-L.; Bacri,~L.; Betton,~J.-M.; Lairez,~D.;
  Pelta,~J.; Auvray,~L. Unfolding of Proteins and Long Transient Conformations
  Detected by Single Nanopore Recording. \emph{Phys. Rev. Lett.} \textbf{2007},
  \emph{98}\relax
\mciteBstWouldAddEndPuncttrue
\mciteSetBstMidEndSepPunct{\mcitedefaultmidpunct}
{\mcitedefaultendpunct}{\mcitedefaultseppunct}\relax
\EndOfBibitem
\bibitem[Oukhaled \latin{et~al.}(2008)Oukhaled, Bacri, Math\'e, Pelta, and
  Auvray]{Oukhaled2008}
Oukhaled,~G.; Bacri,~L.; Math\'e,~J.; Pelta,~J.; Auvray,~L. Effect of Screening
  on the Transport of Polyelectrolytes through Nanopores. \emph{EPL}
  \textbf{2008}, \emph{82}, 48003\relax
\mciteBstWouldAddEndPuncttrue
\mciteSetBstMidEndSepPunct{\mcitedefaultmidpunct}
{\mcitedefaultendpunct}{\mcitedefaultseppunct}\relax
\EndOfBibitem
\bibitem[Stoddart \latin{et~al.}(2009)Stoddart, Heron, Mikhailova, Maglia, and
  Bayley]{Stoddart2009}
Stoddart,~D.; Heron,~A.~J.; Mikhailova,~E.; Maglia,~G.; Bayley,~H.
  Single-Nucleotide Discrimination in Immobilized {DNA} Oligonucleotides with a
  Biological Nanopore. \emph{Proc. Natl. Acad. Sci. U. S. A.} \textbf{2009},
  \emph{106}, 7702--7707\relax
\mciteBstWouldAddEndPuncttrue
\mciteSetBstMidEndSepPunct{\mcitedefaultmidpunct}
{\mcitedefaultendpunct}{\mcitedefaultseppunct}\relax
\EndOfBibitem
\bibitem[Stoddart \latin{et~al.}(2010)Stoddart, Heron, Klingelhoefer,
  Mikhailova, Maglia, and Bayley]{Stoddart2010}
Stoddart,~D.; Heron,~A.~J.; Klingelhoefer,~J.; Mikhailova,~E.; Maglia,~G.;
  Bayley,~H. Nucleobase Recognition in ssDNA at the Central Constriction of the
  $\alpha$-Hemolysin Pore. \emph{Nano Lett.} \textbf{2010}, \emph{10},
  3633--3637\relax
\mciteBstWouldAddEndPuncttrue
\mciteSetBstMidEndSepPunct{\mcitedefaultmidpunct}
{\mcitedefaultendpunct}{\mcitedefaultseppunct}\relax
\EndOfBibitem
\bibitem[Deamer \latin{et~al.}(2016)Deamer, Akeson, and Branton]{Deamer2016}
Deamer,~D.; Akeson,~M.; Branton,~D. {Three Decades of Nanopore Sequencing}.
  \emph{Nat. Biotechnol.} \textbf{2016}, \emph{34}, 518\relax
\mciteBstWouldAddEndPuncttrue
\mciteSetBstMidEndSepPunct{\mcitedefaultmidpunct}
{\mcitedefaultendpunct}{\mcitedefaultseppunct}\relax
\EndOfBibitem
\bibitem[Mathé \latin{et~al.}(2004)Mathé, Visram, Viasnoff, Rabin, and
  Meller]{Mathe2004}
Mathé,~J.; Visram,~H.; Viasnoff,~V.; Rabin,~Y.; Meller,~A. Nanopore Unzipping
  of Individual DNA Hairpin Molecules. \emph{Biophys. J.} \textbf{2004},
  \emph{87}, 3205--3212\relax
\mciteBstWouldAddEndPuncttrue
\mciteSetBstMidEndSepPunct{\mcitedefaultmidpunct}
{\mcitedefaultendpunct}{\mcitedefaultseppunct}\relax
\EndOfBibitem
\bibitem[Mathé \latin{et~al.}(2005)Mathé, Aksimentiev, Nelson, Schulten, and
  Meller]{Mathe2005}
Mathé,~J.; Aksimentiev,~A.; Nelson,~D.~R.; Schulten,~K.; Meller,~A.
  Orientation Discrimination of Single-Stranded DNA Inside the
  $\alpha$-Hemolysin Membrane Channel. \emph{Proc. Natl. Acad. Sci. U. S. A.}
  \textbf{2005}, \emph{102}, 12377--12382\relax
\mciteBstWouldAddEndPuncttrue
\mciteSetBstMidEndSepPunct{\mcitedefaultmidpunct}
{\mcitedefaultendpunct}{\mcitedefaultseppunct}\relax
\EndOfBibitem
\bibitem[Muzard \latin{et~al.}(2010)Muzard, Martinho, Mathé, Bockelmann, and
  Viasnoff]{Muzard2010}
Muzard,~J.; Martinho,~M.; Mathé,~J.; Bockelmann,~U.; Viasnoff,~V. DNA
  Translocation and Unzipping through a Nanopore: Some Geometrical Effects.
  \emph{Biophys. J.} \textbf{2010}, \emph{98}, 2170--2178\relax
\mciteBstWouldAddEndPuncttrue
\mciteSetBstMidEndSepPunct{\mcitedefaultmidpunct}
{\mcitedefaultendpunct}{\mcitedefaultseppunct}\relax
\EndOfBibitem
\bibitem[Di~Marino \latin{et~al.}(2015)Di~Marino, Bonome, Tramontano, and
  Chinappi]{DiMarino2015}
Di~Marino,~D.; Bonome,~E.~L.; Tramontano,~A.; Chinappi,~M. All-Atom Molecular
  Dynamics Simulation of Protein Translocation through an $\alpha$-Hemolysin
  Nanopore. \emph{J. Phys. Chem. Lett.} \textbf{2015}, \emph{6},
  2963--2968\relax
\mciteBstWouldAddEndPuncttrue
\mciteSetBstMidEndSepPunct{\mcitedefaultmidpunct}
{\mcitedefaultendpunct}{\mcitedefaultseppunct}\relax
\EndOfBibitem
\bibitem[Di~Muccio \latin{et~al.}(2019)Di~Muccio, Rossini, Di~Marino, Zollo,
  and Chinappi]{DiMuccio2019}
Di~Muccio,~G.; Rossini,~A.~E.; Di~Marino,~D.; Zollo,~G.; Chinappi,~M. Insights
  into Protein Sequencing with an $\alpha$-{Hemolysin} Nanopore by Atomistic
  Simulations. \emph{Sci. Rep.} \textbf{2019}, \emph{9}\relax
\mciteBstWouldAddEndPuncttrue
\mciteSetBstMidEndSepPunct{\mcitedefaultmidpunct}
{\mcitedefaultendpunct}{\mcitedefaultseppunct}\relax
\EndOfBibitem
\bibitem[Bonome \latin{et~al.}(2019)Bonome, Cecconi, and Chinappi]{Bonome2019}
Bonome,~E.~L.; Cecconi,~F.; Chinappi,~M. Translocation Intermediates of
  Ubiquitin through an $\alpha$-Hemolysin Nanopore: Implications for Detection
  of Post-Translational Modifications. \emph{Nanoscale} \textbf{2019},
  \emph{11}, 9920--9930\relax
\mciteBstWouldAddEndPuncttrue
\mciteSetBstMidEndSepPunct{\mcitedefaultmidpunct}
{\mcitedefaultendpunct}{\mcitedefaultseppunct}\relax
\EndOfBibitem
\bibitem[Fennouri \latin{et~al.}(2018)Fennouri, Ramiandrisoa, Bacri, and
  Math\'e]{Fennouri2018}
Fennouri,~A.; Ramiandrisoa,~J.; Bacri,~L.; Math\'e,~R.,~J\'er\^ome ans~Danie
  Comparative Biosensing of Glycosaminoglycan Hyaluronic Acid Oligo- and
  Polysaccharides Using Aerolysin and $\alpha$-Hemolysin Nanopores. \emph{Eur.
  Phys. J. E} \textbf{2018}, \emph{41}, 1--7\relax
\mciteBstWouldAddEndPuncttrue
\mciteSetBstMidEndSepPunct{\mcitedefaultmidpunct}
{\mcitedefaultendpunct}{\mcitedefaultseppunct}\relax
\EndOfBibitem
\bibitem[Krasilnikov \latin{et~al.}(2006)Krasilnikov, Rodrigues, and
  Bezrukov]{Krasilnikov2006}
Krasilnikov,~O.~V.; Rodrigues,~C.~G.; Bezrukov,~S.~M. Single Polymer Molecules
  in a Protein Nanopore in the Limit of a Strong Polymer-Pore Attraction.
  \emph{Phys. Rev. Lett.} \textbf{2006}, \emph{97}, 018301\relax
\mciteBstWouldAddEndPuncttrue
\mciteSetBstMidEndSepPunct{\mcitedefaultmidpunct}
{\mcitedefaultendpunct}{\mcitedefaultseppunct}\relax
\EndOfBibitem
\bibitem[Reiner \latin{et~al.}(2010)Reiner, Kasianowicz, Nablo, and
  Robertson]{Reiner2010}
Reiner,~J.~E.; Kasianowicz,~J.~J.; Nablo,~B.~J.; Robertson,~J. W.~F. Theory for
  Polymer Analysis Using Nanopore-Based Single-Molecule Mass Spectrometry.
  \emph{Proc. Natl. Acad. Sci. U. S. A.} \textbf{2010}, \emph{107},
  12080--12085\relax
\mciteBstWouldAddEndPuncttrue
\mciteSetBstMidEndSepPunct{\mcitedefaultmidpunct}
{\mcitedefaultendpunct}{\mcitedefaultseppunct}\relax
\EndOfBibitem
\bibitem[Merzlyak \latin{et~al.}(2005)Merzlyak, Capistrano, Valeva,
  Kasianowicz, and Krasilnikov]{Merzlyak2005}
Merzlyak,~P.~G.; Capistrano,~M.-F.~P.; Valeva,~A.; Kasianowicz,~J.~J.;
  Krasilnikov,~O.~V. Conductance and Ion Selectivity of a Mesoscopic Protein
  Nanopore Probed with Cysteine Scanning Mutagenesis. \emph{Biophys. J.}
  \textbf{2005}, \emph{89}, 3059--3070\relax
\mciteBstWouldAddEndPuncttrue
\mciteSetBstMidEndSepPunct{\mcitedefaultmidpunct}
{\mcitedefaultendpunct}{\mcitedefaultseppunct}\relax
\EndOfBibitem
\bibitem[Piguet \latin{et~al.}(2014)Piguet, Discala, Breton, Pelta, Bacri, and
  Oukhaled]{Piguet2014}
Piguet,~F.; Discala,~F.; Breton,~M.-F.; Pelta,~J.; Bacri,~L.; Oukhaled,~A.
  Electroosmosis through $\alpha$-Hemolysin That Depends on Alkali Cation Type.
  \emph{J. Phys. Chem. Lett.} \textbf{2014}, \emph{5}, 4362--4367\relax
\mciteBstWouldAddEndPuncttrue
\mciteSetBstMidEndSepPunct{\mcitedefaultmidpunct}
{\mcitedefaultendpunct}{\mcitedefaultseppunct}\relax
\EndOfBibitem
\bibitem[Payet \latin{et~al.}(2015)Payet, Martinho, Merstorf,
  Pastoriza-Gallego, Pelta, Viasnoff, Auvray, Muthukumar, and
  Mathé]{Payet2015}
Payet,~L.; Martinho,~M.; Merstorf,~C.; Pastoriza-Gallego,~M.; Pelta,~J.;
  Viasnoff,~V.; Auvray,~L.; Muthukumar,~M.; Mathé,~J. Temperature Effect on
  Ionic Current and ssDNA Transport through Nanopores. \emph{Biophys. J.}
  \textbf{2015}, \emph{109}, 1600--1607\relax
\mciteBstWouldAddEndPuncttrue
\mciteSetBstMidEndSepPunct{\mcitedefaultmidpunct}
{\mcitedefaultendpunct}{\mcitedefaultseppunct}\relax
\EndOfBibitem
\bibitem[Misakian and Kasianowicz(2003)Misakian, and Kasianowicz]{Misakian2003}
Misakian,~M.; Kasianowicz,~J.~J. Electrostatic Influence on Ion Transport
  through the {aHL} Channel. \emph{J. Membr. Biol.} \textbf{2003}, \emph{195},
  137--146\relax
\mciteBstWouldAddEndPuncttrue
\mciteSetBstMidEndSepPunct{\mcitedefaultmidpunct}
{\mcitedefaultendpunct}{\mcitedefaultseppunct}\relax
\EndOfBibitem
\bibitem[Noskov \latin{et~al.}(2004)Noskov, Im, and Roux]{Noskov2004}
Noskov,~S.~Y.; Im,~W.; Roux,~B. Ion Permeation through the $\alpha$-Hemolysin
  Channel: Theoretical Studies Based on {Brownian} Dynamics and
  {Poisson}-{Nernst}-{Plank} Electrodiffusion Theory. \emph{Biophys. J.}
  \textbf{2004}, \emph{87}, 2299--2309\relax
\mciteBstWouldAddEndPuncttrue
\mciteSetBstMidEndSepPunct{\mcitedefaultmidpunct}
{\mcitedefaultendpunct}{\mcitedefaultseppunct}\relax
\EndOfBibitem
\bibitem[Aksimentiev and Schulten(2005)Aksimentiev, and
  Schulten]{Aksimentiev2005}
Aksimentiev,~A.; Schulten,~K. Imaging $\alpha$-Hemolysin with Molecular
  Dynamics: Ionic Conductance, Osmotic Permeability, and the Electrostatic
  Potential Map. \emph{Biophys. J.} \textbf{2005}, \emph{88}, 3745--3761\relax
\mciteBstWouldAddEndPuncttrue
\mciteSetBstMidEndSepPunct{\mcitedefaultmidpunct}
{\mcitedefaultendpunct}{\mcitedefaultseppunct}\relax
\EndOfBibitem
\bibitem[Millar \latin{et~al.}(2008)Millar, Madathil, Beckstein, Sansom, Roy,
  and Asenov]{Millar2008}
Millar,~C.; Madathil,~R.; Beckstein,~O.; Sansom,~M. S.~P.; Roy,~S.; Asenov,~A.
  Brownian Simulation of Charge Transport in $\alpha$-{Haemolysin}. \emph{J.
  Comput. Electron.} \textbf{2008}, \emph{7}, 28--33\relax
\mciteBstWouldAddEndPuncttrue
\mciteSetBstMidEndSepPunct{\mcitedefaultmidpunct}
{\mcitedefaultendpunct}{\mcitedefaultseppunct}\relax
\EndOfBibitem
\bibitem[Bhattacharya \latin{et~al.}(2011)Bhattacharya, Muzard, Payet, Mathé,
  Bockelmann, Aksimentiev, and Viasnoff]{Bhattacharya2011}
Bhattacharya,~S.; Muzard,~J.; Payet,~L.; Mathé,~J.; Bockelmann,~U.;
  Aksimentiev,~A.; Viasnoff,~V. Rectification of the Current in
  $\alpha$-Hemolysin Pore Depends on the Cation Type: The Alkali Series Probed
  by Molecular Dynamics Simulations and Experiments. \emph{J. Phys. Chem. C}
  \textbf{2011}, \emph{115}, 4255--4264\relax
\mciteBstWouldAddEndPuncttrue
\mciteSetBstMidEndSepPunct{\mcitedefaultmidpunct}
{\mcitedefaultendpunct}{\mcitedefaultseppunct}\relax
\EndOfBibitem
\bibitem[Gamble \latin{et~al.}(2014)Gamble, Decker, Plett, Pevarnik,
  Pietschmann, Vlassiouk, Aksimentiev, and Siwy]{Gamble2014}
Gamble,~T.; Decker,~K.; Plett,~T.~S.; Pevarnik,~M.; Pietschmann,~J.-F.;
  Vlassiouk,~I.; Aksimentiev,~A.; Siwy,~Z.~S. Rectification of Ion Current in
  Nanopores Depends on the Type of Monovalent Cations: Experiments and
  Modeling. \emph{J. Phys. Chem. C} \textbf{2014}, \emph{118}, 9809--9819\relax
\mciteBstWouldAddEndPuncttrue
\mciteSetBstMidEndSepPunct{\mcitedefaultmidpunct}
{\mcitedefaultendpunct}{\mcitedefaultseppunct}\relax
\EndOfBibitem
\bibitem[Manara \latin{et~al.}(2015)Manara, Guy, Wallace, and
  Khalid]{Manara2015}
Manara,~R. M.~A.; Guy,~A.~T.; Wallace,~E.~J.; Khalid,~S. Free-Energy
  Calculations Reveal the Subtle Differences in the Interactions of DNA Bases
  with $\alpha$-Hemolysin. \emph{J. Chem. Theory Comput.} \textbf{2015},
  \emph{11}, 810--816, PMID: 26579606\relax
\mciteBstWouldAddEndPuncttrue
\mciteSetBstMidEndSepPunct{\mcitedefaultmidpunct}
{\mcitedefaultendpunct}{\mcitedefaultseppunct}\relax
\EndOfBibitem
\bibitem[Bonome \latin{et~al.}(2017)Bonome, Cecconi, and Chinappi]{Bonome2017}
Bonome,~E.~L.; Cecconi,~F.; Chinappi,~M. Electroosmotic Flow through an
  $\alpha$-Hemolysin Nanopore. \emph{Microfluid. Nanofluid.} \textbf{2017},
  \emph{21}\relax
\mciteBstWouldAddEndPuncttrue
\mciteSetBstMidEndSepPunct{\mcitedefaultmidpunct}
{\mcitedefaultendpunct}{\mcitedefaultseppunct}\relax
\EndOfBibitem
\bibitem[Zhou \latin{et~al.}(2020)Zhou, Qiu, Guo, and Guo]{Zhou2020}
Zhou,~W.; Qiu,~H.; Guo,~Y.; Guo,~W. Molecular Insights into Distinct Detection
  Properties of $\alpha$-Hemolysin, MspA, CsgG, and Aerolysin Nanopore Sensors.
  \emph{J. Phys. Chem. B} \textbf{2020}, \emph{124}, 1611--1618, PMID:
  32027510\relax
\mciteBstWouldAddEndPuncttrue
\mciteSetBstMidEndSepPunct{\mcitedefaultmidpunct}
{\mcitedefaultendpunct}{\mcitedefaultseppunct}\relax
\EndOfBibitem
\bibitem[Menestrina(1986)]{Menestrina1986}
Menestrina,~G. Ionic Channels Formed by {Staphylococcus} {aureus} Alpha-Toxin:
  Voltage-Dependent Inhibition by Divalent and Trivalent Cations. \emph{J.
  Membr. Biol.} \textbf{1986}, \emph{90}, 177--190\relax
\mciteBstWouldAddEndPuncttrue
\mciteSetBstMidEndSepPunct{\mcitedefaultmidpunct}
{\mcitedefaultendpunct}{\mcitedefaultseppunct}\relax
\EndOfBibitem
\bibitem[Bonthuis \latin{et~al.}(2006)Bonthuis, Zhang, Hornblower, Mathé,
  Shklovskii, and Meller]{Bonthuis2006}
Bonthuis,~D.~J.; Zhang,~J.; Hornblower,~B.; Mathé,~J.; Shklovskii,~B.~I.;
  Meller,~A. Self-Energy-Limited Ion Transport in Subnanometer Channels.
  \emph{Phys. Rev. Lett.} \textbf{2006}, \emph{97}, 128104\relax
\mciteBstWouldAddEndPuncttrue
\mciteSetBstMidEndSepPunct{\mcitedefaultmidpunct}
{\mcitedefaultendpunct}{\mcitedefaultseppunct}\relax
\EndOfBibitem
\bibitem[Gu \latin{et~al.}(2000)Gu, Dalla~Serra, Vincent, Vigh, Cheley, Braha,
  and Bayley]{Gu2000}
Gu,~L.-Q.; Dalla~Serra,~M.; Vincent,~J.~B.; Vigh,~G.; Cheley,~S.; Braha,~O.;
  Bayley,~H. Reversal of Charge Selectivity in Transmembrane Protein Pores by
  Using Noncovalent Molecular Adapters. \emph{Proc. Natl. Acad. Sci. U. S. A.}
  \textbf{2000}, \emph{97}, 3959--3964\relax
\mciteBstWouldAddEndPuncttrue
\mciteSetBstMidEndSepPunct{\mcitedefaultmidpunct}
{\mcitedefaultendpunct}{\mcitedefaultseppunct}\relax
\EndOfBibitem
\bibitem[Mohammad and Movileanu(2010)Mohammad, and Movileanu]{Mohammad2010}
Mohammad,~M.~M.; Movileanu,~L. Impact of Distant Charge Reversals within a
  Robust $\beta$-Barrel Protein Pore. \emph{J. Phys. Chem. B} \textbf{2010},
  \emph{114}, 8750--8759\relax
\mciteBstWouldAddEndPuncttrue
\mciteSetBstMidEndSepPunct{\mcitedefaultmidpunct}
{\mcitedefaultendpunct}{\mcitedefaultseppunct}\relax
\EndOfBibitem
\bibitem[Basdevant \latin{et~al.}(2019)Basdevant, Dessaux, and
  Ramirez]{Basdevant2019}
Basdevant,~N.; Dessaux,~D.; Ramirez,~R. Ionic Transport through a Protein
  Nanopore: a Coarse-Grained Molecular Dynamics Study. \emph{Sci. Rep.}
  \textbf{2019}, \emph{9}, 15740\relax
\mciteBstWouldAddEndPuncttrue
\mciteSetBstMidEndSepPunct{\mcitedefaultmidpunct}
{\mcitedefaultendpunct}{\mcitedefaultseppunct}\relax
\EndOfBibitem
\bibitem[Periole \latin{et~al.}(2009)Periole, Cavalli, Marrink, and
  Ceruso]{Periole2009}
Periole,~X.; Cavalli,~M.; Marrink,~S.-J.; Ceruso,~M.~A. Combining an Elastic
  Network With a Coarse-Grained Molecular Force Field: Structure, Dynamics, and
  Intermolecular Recognition. \emph{J. Chem. Theory Comput.} \textbf{2009},
  \emph{5}, 2531--2543\relax
\mciteBstWouldAddEndPuncttrue
\mciteSetBstMidEndSepPunct{\mcitedefaultmidpunct}
{\mcitedefaultendpunct}{\mcitedefaultseppunct}\relax
\EndOfBibitem
\bibitem[Gumbart \latin{et~al.}(2012)Gumbart, Khalili-Araghi, Sotomayor, and
  Roux]{Gumbart2012}
Gumbart,~J.; Khalili-Araghi,~F.; Sotomayor,~M.; Roux,~B. Constant Electric
  Field Simulations of the Membrane Potential Illustrated with Simple Systems.
  \emph{Biochim. Biophys. Acta, Biomembr.} \textbf{2012}, \emph{1818},
  294--302\relax
\mciteBstWouldAddEndPuncttrue
\mciteSetBstMidEndSepPunct{\mcitedefaultmidpunct}
{\mcitedefaultendpunct}{\mcitedefaultseppunct}\relax
\EndOfBibitem
\bibitem[Modi \latin{et~al.}(2012)Modi, Winterhalter, and
  Kleinekathöfer]{Modi2012}
Modi,~N.; Winterhalter,~M.; Kleinekathöfer,~U. Computational Modeling of Ion
  Transport through Nanopores. \emph{Nanoscale} \textbf{2012}, \emph{4},
  6166--6180\relax
\mciteBstWouldAddEndPuncttrue
\mciteSetBstMidEndSepPunct{\mcitedefaultmidpunct}
{\mcitedefaultendpunct}{\mcitedefaultseppunct}\relax
\EndOfBibitem
\bibitem[Aksimentiev(2010)]{Aksimentiev2010}
Aksimentiev,~A. Deciphering Ionic Current Signatures of {DNA} Transport through
  a Nanopore. \emph{Nanoscale} \textbf{2010}, \emph{2}, 468--483\relax
\mciteBstWouldAddEndPuncttrue
\mciteSetBstMidEndSepPunct{\mcitedefaultmidpunct}
{\mcitedefaultendpunct}{\mcitedefaultseppunct}\relax
\EndOfBibitem
\bibitem[Crozier \latin{et~al.}(2001)Crozier, Henderson, Rowley, and
  Busath]{Crozier2001}
Crozier,~P.~S.; Henderson,~D.; Rowley,~R.~L.; Busath,~D.~D. Model Channel Ion
  Currents in NaCl-Extended Simple Point Charge Water Solution with
  Applied-Field Molecular Dynamics. \emph{Biophys. J.} \textbf{2001},
  \emph{81}, 3077--3089\relax
\mciteBstWouldAddEndPuncttrue
\mciteSetBstMidEndSepPunct{\mcitedefaultmidpunct}
{\mcitedefaultendpunct}{\mcitedefaultseppunct}\relax
\EndOfBibitem
\bibitem[Walker \latin{et~al.}(1992)Walker, Krishnasastry, Zorn, Kasianowicz,
  and Bayley]{Walker1992}
Walker,~B.; Krishnasastry,~M.; Zorn,~L.; Kasianowicz,~J.; Bayley,~H. Functional
  Expression of the $\alpha$-Hemolysin of {Staphylococcus} {aureus} in Intact
  {Escherichia} {coli} and in Cell Lysates. Deletion of Five C-Terminal Amino
  Acids Selectively Impairs Hemolytic Activity. \emph{J. Biol. Chem.}
  \textbf{1992}, \emph{267}, 10902--10909\relax
\mciteBstWouldAddEndPuncttrue
\mciteSetBstMidEndSepPunct{\mcitedefaultmidpunct}
{\mcitedefaultendpunct}{\mcitedefaultseppunct}\relax
\EndOfBibitem
\bibitem[Simakov and Kurnikova(2010)Simakov, and Kurnikova]{Simakov2010}
Simakov,~N.~A.; Kurnikova,~M.~G. Soft Wall Ion Channel in Continuum
  Representation with Application to Modeling Ion Currents in
  $\alpha$-Hemolysin. \emph{J. Phys. Chem. B} \textbf{2010}, \emph{114},
  15180--15190\relax
\mciteBstWouldAddEndPuncttrue
\mciteSetBstMidEndSepPunct{\mcitedefaultmidpunct}
{\mcitedefaultendpunct}{\mcitedefaultseppunct}\relax
\EndOfBibitem
\bibitem[Dyrka \latin{et~al.}(2013)Dyrka, Bartuzel, and Kotulska]{Dyrka2013}
Dyrka,~W.; Bartuzel,~M.~M.; Kotulska,~M. Optimization of 3D
  {Poisson}-{Nernst}-{Planck} Model for Fast Evaluation of Diverse Protein
  Channels: 3D {PNP} for Diverse Protein Channels. \emph{Proteins: Struct.,
  Funct., Bioinf.} \textbf{2013}, \emph{81}, 1802--1822\relax
\mciteBstWouldAddEndPuncttrue
\mciteSetBstMidEndSepPunct{\mcitedefaultmidpunct}
{\mcitedefaultendpunct}{\mcitedefaultseppunct}\relax
\EndOfBibitem
\bibitem[Boukhet \latin{et~al.}(2016)Boukhet, Piguet, Ouldali,
  Pastoriza-Gallego, Pelta, and Oukhaled]{Boukhet2016}
Boukhet,~M.; Piguet,~F.; Ouldali,~H.; Pastoriza-Gallego,~M.; Pelta,~J.;
  Oukhaled,~A. Probing Driving Forces in Aerolysin and $\alpha$-Hemolysin
  Biological Nanopores: Electrophoresis Versus Electroosmosis. \emph{Nanoscale}
  \textbf{2016}, \emph{8}, 18352--18359\relax
\mciteBstWouldAddEndPuncttrue
\mciteSetBstMidEndSepPunct{\mcitedefaultmidpunct}
{\mcitedefaultendpunct}{\mcitedefaultseppunct}\relax
\EndOfBibitem
\bibitem[Maglia \latin{et~al.}(2008)Maglia, Restrepo, Mikhailova, and
  Bayley]{Maglia2008}
Maglia,~G.; Restrepo,~M.~R.; Mikhailova,~E.; Bayley,~H. Enhanced Translocation
  of Single DNA Molecules through $\alpha$-Hemolysin Nanopores by Manipulation
  of Internal Charge. \emph{Proc. Natl. Acad. Sci. U. S. A.} \textbf{2008},
  \emph{105}, 19720--19725\relax
\mciteBstWouldAddEndPuncttrue
\mciteSetBstMidEndSepPunct{\mcitedefaultmidpunct}
{\mcitedefaultendpunct}{\mcitedefaultseppunct}\relax
\EndOfBibitem
\bibitem[Rincon-Restrepo \latin{et~al.}(2011)Rincon-Restrepo, Mikhailova,
  Bayley, and Maglia]{Rincon-Restrepo2011}
Rincon-Restrepo,~M.; Mikhailova,~E.; Bayley,~H.; Maglia,~G. Controlled
  Translocation of Individual DNA Molecules through Protein Nanopores with
  Engineered Molecular Brakes. \emph{Nano Lett.} \textbf{2011}, \emph{11},
  746--750\relax
\mciteBstWouldAddEndPuncttrue
\mciteSetBstMidEndSepPunct{\mcitedefaultmidpunct}
{\mcitedefaultendpunct}{\mcitedefaultseppunct}\relax
\EndOfBibitem
\bibitem[Ayub and Bayley(2012)Ayub, and Bayley]{Ayub2012}
Ayub,~M.; Bayley,~H. Individual RNA Base Recognition in Immobilized
  Oligonucleotides Using a Protein Nanopore. \emph{Nano Lett.} \textbf{2012},
  \emph{12}, 5637--5643\relax
\mciteBstWouldAddEndPuncttrue
\mciteSetBstMidEndSepPunct{\mcitedefaultmidpunct}
{\mcitedefaultendpunct}{\mcitedefaultseppunct}\relax
\EndOfBibitem
\bibitem[Berman \latin{et~al.}(2000)Berman, Westbrook, Feng, Gilliland, Bhat,
  Weissig, Shindyalov, and Bourne]{Berman2000}
Berman,~H.~M.; Westbrook,~J.; Feng,~Z.; Gilliland,~G.; Bhat,~T.~N.;
  Weissig,~H.; Shindyalov,~I.~N.; Bourne,~P.~E. The Protein Data Bank.
  \emph{Nucleic Acids Res.} \textbf{2000}, \emph{28}, 235--242\relax
\mciteBstWouldAddEndPuncttrue
\mciteSetBstMidEndSepPunct{\mcitedefaultmidpunct}
{\mcitedefaultendpunct}{\mcitedefaultseppunct}\relax
\EndOfBibitem
\bibitem[Dolinsky \latin{et~al.}(2004)Dolinsky, Nielsen, McCammon, and
  Baker]{Dolinsky2004}
Dolinsky,~T.~J.; Nielsen,~J.~E.; McCammon,~J.~A.; Baker,~N.~A. {PDB2PQR}: an
  Automated Pipeline for the Setup of {Poisson}-{Boltzmann} Electrostatics
  Calculations. \emph{Nucleic Acids Res.} \textbf{2004}, \emph{32},
  W665--W667\relax
\mciteBstWouldAddEndPuncttrue
\mciteSetBstMidEndSepPunct{\mcitedefaultmidpunct}
{\mcitedefaultendpunct}{\mcitedefaultseppunct}\relax
\EndOfBibitem
\bibitem[Olsson \latin{et~al.}(2011)Olsson, Søndergaard, Rostkowski, and
  Jensen]{Olsson2011}
Olsson,~M. H.~M.; Søndergaard,~C.~R.; Rostkowski,~M.; Jensen,~J.~H. {PROPKA3}:
  Consistent Treatment of Internal and Surface Residues in Empirical pKa
  Predictions. \emph{J. Chem. Theory Comput.} \textbf{2011}, \emph{7},
  525--537\relax
\mciteBstWouldAddEndPuncttrue
\mciteSetBstMidEndSepPunct{\mcitedefaultmidpunct}
{\mcitedefaultendpunct}{\mcitedefaultseppunct}\relax
\EndOfBibitem
\bibitem[de~Jong \latin{et~al.}(2013)de~Jong, Singh, Bennett, Arnarez,
  Wassenaar, Schäfer, Periole, Tieleman, and Marrink]{deJong2013}
de~Jong,~D.~H.; Singh,~G.; Bennett,~W. F.~D.; Arnarez,~C.; Wassenaar,~T.~A.;
  Schäfer,~L.~V.; Periole,~X.; Tieleman,~D.~P.; Marrink,~S.~J. Improved
  Parameters for the {Martini} Coarse-Grained Protein Force Field. \emph{J.
  Chem. Theory Comput.} \textbf{2013}, \emph{9}, 687--697\relax
\mciteBstWouldAddEndPuncttrue
\mciteSetBstMidEndSepPunct{\mcitedefaultmidpunct}
{\mcitedefaultendpunct}{\mcitedefaultseppunct}\relax
\EndOfBibitem
\bibitem[Scott \latin{et~al.}(2008)Scott, Bond, Ivetac, Chetwynd, Khalid, and
  Sansom]{Scott2008}
Scott,~K.~A.; Bond,~P.~J.; Ivetac,~A.; Chetwynd,~A.~P.; Khalid,~S.; Sansom,~M.
  S.~P. Coarse-Grained {MD} Simulations of Membrane Protein-Bilayer
  Self-Assembly. \emph{Structure} \textbf{2008}, \emph{16}, 621--630\relax
\mciteBstWouldAddEndPuncttrue
\mciteSetBstMidEndSepPunct{\mcitedefaultmidpunct}
{\mcitedefaultendpunct}{\mcitedefaultseppunct}\relax
\EndOfBibitem
\bibitem[Yesylevskyy \latin{et~al.}(2010)Yesylevskyy, Schäfer, Sengupta, and
  Marrink]{Yesylevskyy2010}
Yesylevskyy,~S.~O.; Schäfer,~L.~V.; Sengupta,~D.; Marrink,~S.~J. Polarizable
  Water Model for the Coarse-Grained {MARTINI} Force Field. \emph{PLoS Comput.
  Biol.} \textbf{2010}, \emph{6}, e1000810\relax
\mciteBstWouldAddEndPuncttrue
\mciteSetBstMidEndSepPunct{\mcitedefaultmidpunct}
{\mcitedefaultendpunct}{\mcitedefaultseppunct}\relax
\EndOfBibitem
\bibitem[Guros \latin{et~al.}(2018)Guros, Balijepalli, and Klauda]{Guros2018}
Guros,~N.~B.; Balijepalli,~A.; Klauda,~J.~B. The Role of Lipid Interactions in
  Simulations of the $\alpha$-Hemolysin Ion-Channel-Forming Toxin.
  \emph{Biophys. J.} \textbf{2018}, \emph{115}, 1720--1730\relax
\mciteBstWouldAddEndPuncttrue
\mciteSetBstMidEndSepPunct{\mcitedefaultmidpunct}
{\mcitedefaultendpunct}{\mcitedefaultseppunct}\relax
\EndOfBibitem
\end{mcitethebibliography}

\end{document}